\documentclass[useAMS,usenatbib]{mn2e}
\usepackage[utf8]{inputenc}
\usepackage{graphicx}
\usepackage{amsmath}
\usepackage{amssymb}
\usepackage{times}
\usepackage{color}
\usepackage[usenames,dvipsnames,svgnames,table]{xcolor}
\usepackage{xspace}
\usepackage{subcaption}

\usepackage{fixltx2e} 

\definecolor{customhdrcolor}{rgb}{0.0,0.0,0.0}
\definecolor{customcitecolor}{rgb}{0.0,0.5,0.75}
\definecolor{customlinkcolor}{rgb}{0.0,0.5,0.75}

\usepackage[colorlinks=true,linkcolor=customlinkcolor,urlcolor=customlinkcolor,citecolor=customcitecolor,pdftex]{hyperref}

\ifpdf\else\usepackage{graphics}\fi

\newcommand{\degree}{\ensuremath{^{\circ}}\xspace}

\newcommand{\bwimgextension}{}

\DeclareRobustCommand{\TUSSEN}[3]{#2}

\title[WSClean: a fast, generic wide-field imager]{WSClean: an implementation of a fast, generic wide-field imager for radio astronomy}

\def\ANU{$^{1}$}
\def\CAASTRO{$^{2}$}
\def\Curtin{$^{3}$}
\def\UWisc{$^{4}$}
\def\USydney{$^{5}$}
\def\MIT{$^{6}$}
\def\SKASA{$^{7}$}
\def\ASU{$^{8}$}
\def\Haystack{$^{9}$}
\def\RRI{$^{10}$}
\def\CfA{$^{11}$}
\def\UW{$^{12}$}
\def\Victoria{$^{13}$}
\def\UMichigan{$^{14}$}
\def\CASS{$^{15}$}
\def\Tata{$^{16}$}
\def\UMelbourne{$^{17}$}
\def\NRAO{$^{18}$}

\author[A.~R.~Offringa et al.]{A.~R.~Offringa$^{1,2}$\thanks{Corresponding author. E-mail: \url{andre.offringa@anu.edu.au}},
B.~McKinley\ANU$^,$\CAASTRO,
N.~Hurley-Walker\Curtin,
F.~H.~Briggs$^{1,2}$,
R.~B.~Wayth\Curtin$^,$\CAASTRO, \newauthor
D.~L.~Kaplan\UWisc,
M.~E.~Bell\USydney$^,$\CAASTRO,
L.~Feng\MIT,
A.~R.~Neben\MIT,
J.~D.~Hughes\ANU, 
J.~Rhee\ANU$^,$\CAASTRO, \newauthor
T.~Murphy\USydney$^,$\CAASTRO,
N.~D.~R.~Bhat\Curtin,
G.~Bernardi\SKASA,
J.~D.~Bowman\ASU,
R.~J.~Cappallo\Haystack,
B.~E.~Corey\Haystack, \newauthor
A.~A.~Deshpande\RRI, 
D.~Emrich\Curtin,
A.~Ewall-Wice\MIT,
B.~M.~Gaensler\USydney$^,$\CAASTRO,
R.~Goeke\MIT, \newauthor
L.~J.~Greenhill\CfA, 
B.~J.~Hazelton\UW, 
L.~Hindson\Victoria,
M.~Johnston-Hollitt\Victoria,
D.~C.~Jacobs\ASU, \newauthor
J.~C.~Kasper\UMichigan$^,$\CfA,
E.~Kratzenberg\Haystack, 
E.~Lenc\USydney$^,$\CAASTRO,
C.~J.~Lonsdale\Haystack,
M.~J.~Lynch\Curtin, \newauthor
S.~R.~McWhirter\Haystack,
D.~A.~Mitchell\CASS$^,$\CAASTRO,
M.~F.~Morales\UW, 
E.~Morgan\MIT,
N.~Kudryavtseva\Curtin, \newauthor
D.~Oberoi\Tata,
S.~M.~Ord\Curtin$^,$\CAASTRO,
B.~Pindor\UMelbourne,
P.~Procopio\UMelbourne,
T.~Prabu\RRI,
J.~Riding\UMelbourne,  \newauthor
D.~A.~Roshi\NRAO,
N.~Udaya~Shankar\RRI,
K.~S.~Srivani\RRI,
R.~Subrahmanyan\RRI$^,$\CAASTRO,
S.~J.~Tingay\Curtin$^,$\CAASTRO, \newauthor
M.~Waterson\Curtin$^,$\ANU,
R.~L.~Webster\UMelbourne$^,$\CAASTRO, 
A.~R.~Whitney\Haystack, 
A.~Williams\Curtin,
C.~L.~Williams\MIT
\\
\ANU{}Research School of Astronomy and Astrophysics, Australian National University, Canberra, ACT 2611, Australia \\
\CAASTRO{}ARC Centre of Excellence for All-sky Astrophysics (CAASTRO), Australian National University, Canberra, ACT 2611, Australia \\
\Curtin{}International Centre for Radio Astronomy Research, Curtin University, Bentley, WA 6102, Australia\\
\UWisc{}Department of Physics, University of Wisconsin--Milwaukee, Milwaukee, WI 53201, USA\\
\USydney{}Sydney Institute for Astronomy, School of Physics, The University of Sydney, NSW 2006, Australia\\
\MIT{}Kavli Institute for Astrophysics and Space Research, Massachusetts Institute of Technology, Cambridge, MA 02139, USA\\
\SKASA{}Square Kilometre Array South Africa (SKA SA), Cape Town 7405, South Africa\\
\ASU{}School of Earth and Space Exploration, Arizona State University, Tempe, AZ 85287, USA\\
\Haystack{}MIT Haystack Observatory, Westford, MA 01886, USA\\
\RRI{}Raman Research Institute, Bangalore 560080, India\\
\CfA{}Harvard-Smithsonian Center for Astrophysics, Cambridge, MA 02138, USA\\
\UW{}Department of Physics, University of Washington, Seattle, WA 98195, USA\\
\Victoria{}School of Chemical \& Physical Sciences, Victoria University of Wellington, Wellington 6140, New Zealand\\
\UMichigan{}University of Michigan, Ann Harbor, MI 48109, USA\\
\CASS{}CSIRO Astronomy and Space Science, Marsfield, NSW 2122, Australia\\
\Tata{}National Centre for Radio Astrophysics, Tata Institute for Fundamental Research, Pune 411007, India\\
\UMelbourne{}School of Physics, The University of Melbourne, Parkville, VIC 3010, Australia\\
\NRAO{}National Radio Astronomy Observatory, Charlottesville and Greenbank, USA
}

\begin{document}

\date{Accepted 2014 July 6.  Received 2014 July 4; in original form 2014 April 1.}
\pagerange{\pageref{firstpage}--\pageref{lastpage}}
\pubyear{2014}

\label{firstpage}
\maketitle

\begin{abstract}
Astronomical widefield imaging of interferometric radio data is computationally expensive, especially for the large data volumes created by modern non-coplanar many-element arrays. We present a new widefield interferometric imager that uses the $w$-stacking algorithm and can make use of the $w$-snapshot algorithm. The performance dependencies of \textsc{casa}'s $w$-projection and our new imager are analysed and analytical functions are derived that describe the required computing cost for both imagers. On data from the Murchison Widefield Array, we find our new method to be an order of magnitude faster than $w$-projection, as well as being capable of full-sky imaging at full resolution and with correct polarisation correction.  We predict the computing costs for several other arrays and estimate that our imager is a factor of 2--12 faster, depending on the array configuration. We estimate the computing cost for imaging the low-frequency Square-Kilometre Array observations to be 60 PetaFLOPS with current techniques. We find that combining $w$-stacking with the $w$-snapshot algorithm does not significantly improve computing requirements over pure $w$-stacking. The source code of our new imager is publicly released.
\end{abstract}


\section{Introduction}
Visibility data from non-coplanar interferometric radio telescopes that observe large fractions of the sky at once can not be accurately imaged with a two-dimensional fast Fourier transform (FFT). Instead, the imaging algorithm needs to account for the ``$w$-term'' during inversion, which is the term that describes the deviation of the array from a perfect plane \citep{perley-noncoplanar-arrays}. The image degradation effects of the $w$-term are amplified for telescopes with wide fields of view (FOV), making this a significant issue for low-frequency telescopes that by nature are wide-field instruments.

There are several methods to deal with the $w$-term during imaging: faceting \citep{facetting-cornwell}; a three-dimensional Fourier transform \citep{perley-noncoplanar-arrays}; $w$-projection \citep*{wprojection-cornwell}; $w$-stacking \citep{ska-memo-regridding-2011}; and warped snapshots \citep{perley-noncoplanar-arrays}. Hybrid methods are sometimes useful, such as with the $w$-snapshots method \citep*{widefield-imaging-ska-cornwell}.

A new generation of wide-field observatories is producing data sets that are orders of magnitude larger than before. Examples of such telescopes include the Murchison Widefield Array (MWA, \citealt{mwa-design-2009}, \citealt{mwa}), the upgraded Jansky Very Large Array (JVLA) and the Low-Frequency Array (LOFAR, \citealt{lofar-2013}). The Common Astronomy Software Applications (\textsc{casa}, \citealt{casa-2007}, \citealt{casa-2008}) have an efficient implementation of the $w$-projection algorithm, with many available features such as multi-scale clean and spectral-shape fitting during deconvolution. However, with the MWA we have seen that imaging a 2-min snapshot observation away from zenith can take up to tens of wall-clock hours with \textsc{casa}'s $w$-projection algorithm, because of the larger $w$-terms for off-zenith observations. Imaging with larger image sizes or at higher zenith angles can be impossible because the size and number of $w$-kernels become too large to hold in memory.

Another option exists for imaging MWA data: the Real-Time System (RTS, \citealt{rts-mwa,mwa-interferometric-imaging}). This has been designed as an efficient calibration and imaging pipeline specifically for MWA data. It can use GPUs to improve efficiency. Snapshot imaging is performed to deal with the $w$-term, which implies that slight variations in tile elevation cause some decorrelation on the longer baselines. As the RTS was designed as a single-pass stream processor, standard iterative deconvolution algorithms are not available. Compact emission can be subtracted and peeled from visibilities using a sky model and calibration updates, but updates to the sky model need to be realised using separate forward-modelling routines \citep{forward-modelling-bernardi-2011, pindor-subtracting-sources-2011}.

To reach high dynamic ranges, it can be necessary to deal with direction-dependent effects (DDEs). This is especially true for wide-field telescopes. One way to correct for known DDEs is by using the $a$-projection technique, which convolves the data during gridding with a kernel that corrects the DDEs \citep{aprojection-2008}. One particular DDE is the effect of the ionosphere. For the MWA it can be assumed that the ionosphere has the same effect on all antennas, because the maximum baseline length is relatively small (2.9~km) and smaller than the typical size of ionospheric structure \citep{lonsdale-calibration-approaches-ionosphere}. This is not the case for LOFAR, making it necessary to correct the direction-dependent ionospheric effects per station before gridding the data. The \textsc{awimager} \citep{awimager-2013} has been written to perform these corrections, and uses a hybrid of a-projection, $w$-projection and $w$-stacking. The $a$-projection technique can only be applied directly for deterministic effects such as the correction of the primary beam. Effects like the ionosphere require separate calibration or estimation before $a$-projection can be applied.

Once the Square-Kilometre Array (SKA) begins its operation, the required computational power for wide-field imaging will become an even bigger challenge. \citet{widefield-imaging-ska-cornwell} argues that the $w$-snapshots algorithm is the most efficient approach for the SKA.

In this article, we present a new implementation of a generic wide-field imager that is significantly faster than \textsc{casa}'s $w$-projection implementation. To obtain the increase in speed, the implementation uses the $w$-stacking method for correcting the $w$-terms, optionally combined with a new technique for $w$-snapshot imaging. We named the new imager ``WSClean'', as an abbreviation for ``$w$-Stacking Clean''. Our new imaging implementation, which in our experience is anywhere from 2 to 12 times faster than the CASA $w$-projection imager, is publicly released\footnote{The \textsc{wsclean} source code can be found at:\\\href{http://sourceforge.net/p/wsclean}{http://sourceforge.net/p/wsclean}}.

This paper is structured as follows: The $w$-stacking algorithm is described in Sect.~\ref{sec:wstacking}. Details of implementing the $w$-stacking and $w$-snapshots algorithms are described in Sect.~\ref{sec:implementation}. The performance and accuracy will be analysed in Sect.~\ref{sec:analysis}. The conclusions are presented in Sect.~\ref{sec:conclusions}.

\section{The $w$-stacking technique} \label{sec:wstacking}
In this section, we will describe the $w$-stacking algorithm from a mathematical point of view. Instead of applying a convolution in $uv$-space, the $w$-stacking method grids visibilities on different $w$-layers and performs the $w$-corrections after the inverse Fourier transforms \citep{ska-memo-regridding-2011}.

An interferometer samples the complex visibility function
\begin{align}\notag
V(u,v,w) = & \iint \frac{A(l,m) I(l,m)}{\sqrt{1-l^2-m^2}} \times \\ \label{eq:visibility-function}
& e^{-2\pi i \left(ul + vm + w(\sqrt{1-l^2-m^2}-1)\right)} dl dm,
\end{align}
where $u,v,w$ is a baseline coordinate in the coordinate system of the array, $A$ is the primary-beam function, $I$ is the sky function and $l,m$ are cosine sky coordinates. We will use $I'(l,m)$ to denote the sky function before primary-beam correction, $I'(l,m)=A(l,m)I(l,m)$. We will not discuss calibration, but assume $V$ has been calibrated before imaging. In the case of a polarised measurement, the symbols become $2\times 2$ matrices and beam correction is more complicated, but without loss of generality we will ignore polarisation and treat inversion as a scalar problem. Imaging consists of inverting Eq.~\eqref{eq:visibility-function}, i.e., to find $I'$ from $V$.

For small FOVs, the term $\sqrt{1-l^2-m^2}$ is approximately of unit size, making Eq.~\eqref{eq:visibility-function} approximately an ordinary invertable two-dimensional Fourier transform. A common rule is that this is valid when
\begin{equation}\label{eq:when-2d-is-valid}
\forall w,l,m: w\left(\sqrt{1-l^2-m^2}-1\right) \ll 1.
\end{equation}

To derive the $w$-stacking technique, Eq.~\eqref{eq:visibility-function} is rewritten to
\begin{align}\notag
V(u,v,w) = & \iint \frac{I'(l,m) e^{-2\pi i w(\sqrt{1-l^2-m^2}-1)}}{\sqrt{1-l^2-m^2}} \times \\ \notag
& e^{-2\pi i \left(ul + vm\right)} dl dm.
\end{align}
This is an ordinary two-dimensional Fourier transform going from $u,v$ space to $l,m$ space, and can be inverted to get:
\begin{align}\notag
\frac{I'(l,m)}{\sqrt{1-l^2-m^2}} = & e^{2\pi i w(\sqrt{1-l^2-m^2}-1)} \iint V(u,v,w) \times \\ \notag
& e^{2\pi i \left(ul + vm\right)} du dv.
\end{align}
Integrating both sides over $w_{\min}$ to $w_{\max}$, the minimum and maximum value of $w$, results in
\begin{align}\notag
\frac{I'(l,m)\left(w_{\max} - w_{\min}\right)}{\sqrt{1-l^2-m^2}} = \int\limits_{w_{\min}}^{w_{\max}} e^{2\pi i w(\sqrt{1-l^2-m^2}-1)} \times \\ \label{eq:wstacking}
\iint V(u,v,w)  e^{2\pi i \left(ul + vm\right)} du dv dw.
\end{align}
The final step is to make the $u,v,w$ parameters discrete, so that the integration over $u$ and $v$ can become an inverse FFT and the integration over $w$ becomes a summation. This shows that the sky function can be reconstructed by: i)~gridding samples with equal $w$-value on a uniform grid; ii)~calculating the inverse FFT; iii)~applying the direction-dependent phase shift $e^{2\pi i w(\sqrt{1-l^2-m^2}-1)}$; iv)~repeating this for all $w$-values and adding the results together; v)~applying the final scaling.

In practice, the final scaling will be different from $\left(w_{\max} - w_{\min}\right)/\sqrt{1-l^2-m^2}$ suggested by Eq.~\eqref{eq:wstacking}, because the individual $w$-layers will not be completely filled with samples. Therefore, each pixel is divided by the weighted number of samples. Additionally, it might be required to divide out the effect of a possible convolution kernel, the primary beam and correct for other direction-dependent effects. In equally-polarized baselines (e.g., XX, YY, LL or RR), a correlated baseline is the complex conjugate of the reversed baseline, and the relation $V(u,v,w)=\overline{V(-u,-v,-w)}$ holds. The right-hand side of Eq.\eqref{eq:wstacking} with only positive $w$-value samples then becomes the complex conjugate of the one with only negative $w$-value samples. In this case we can therefore calculate the image for $w<0$ from the image with $w>0$. That allows us to set $w_{\min}$ to the minimum absolute $w$-value, which requires half the number of layers. In any case, the input to the two-dimensional inverse FFT is not generally a Hermitian-symmetric function, and hence the inverse FFT is always performed as a complex-to-complex transform.

The inverse of imaging, i.e., to calculate the visibility from a model image, can be done by reversing the $w$-stacking algorithm: i)~multiply the image with the appropriate factor; ii)~copy the image to several layers; iii)~inverse apply the direction-dependent phase shift for each layer; iv)~FFT each layer; and v)~sample a required visibility from the correct $w$-layer. We will refer to this operation as prediction.

\subsection{Discretisation of $w$} \label{sec:gridding-w}
\begin{figure}
\begin{center}
\includegraphics[width=8cm]{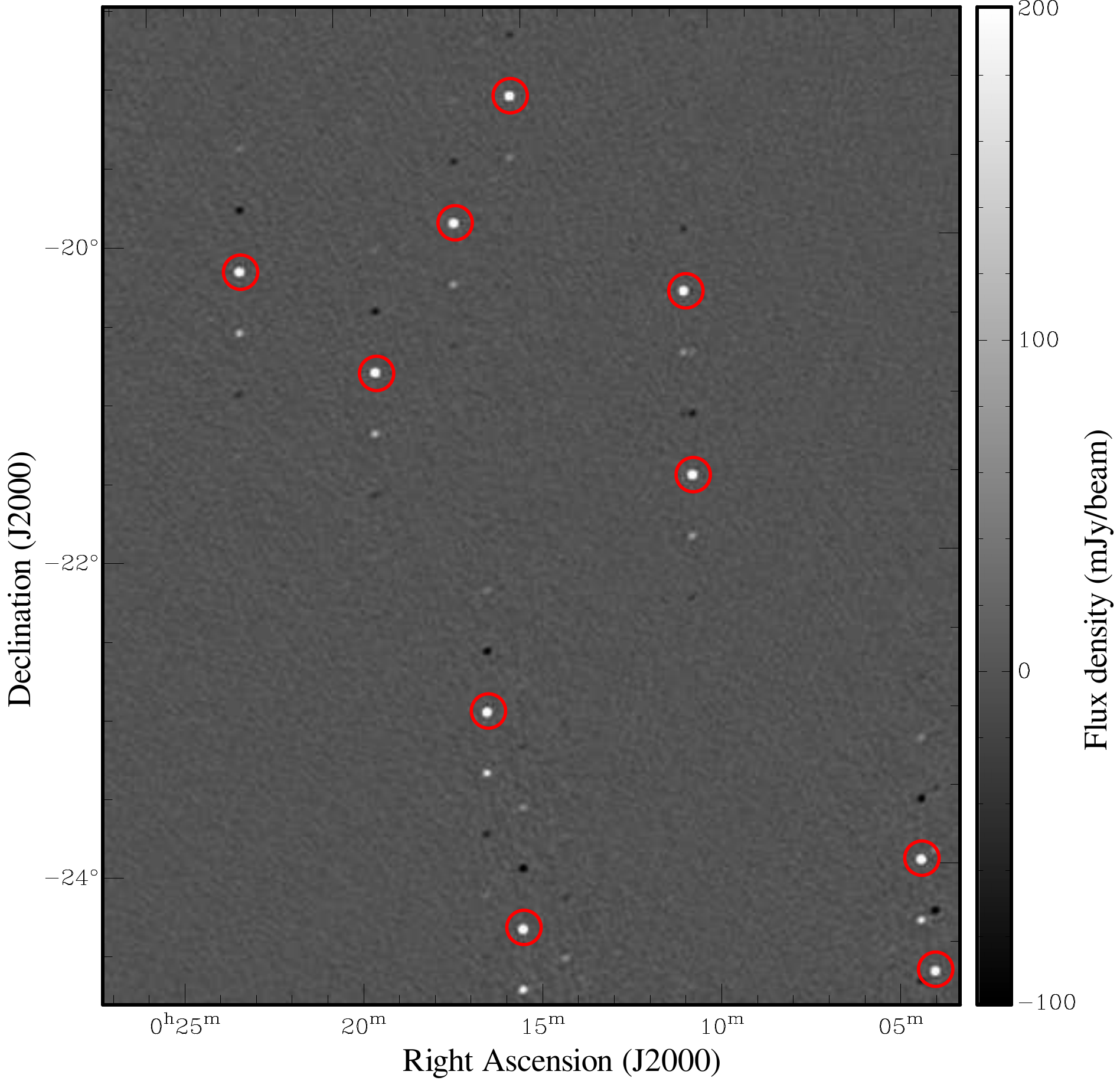}
\caption{Aliasing artefacts caused by insufficient $w$-layers in a simulated field. \textsc{wsclean} was set to use 12 $w$-layers. The centre of the image is at 10\degree zenith angle, which would normally require $\sim195$ $w$-layers. Sources are 1~Jy (red circles), ghost sources are approximately 0.2~Jy. Each source produces two ghost sources, but because they re-appear after a major cleaning cycle, they are eventually cleaned and produce more ghost sources. }
\label{fig:aliasing-example}
\end{center}
\end{figure}
While the discretisation of $u$ and $v$ is similar to conventional imaging, the discretisation of $w$ defines the number of $w$-layers that need to be processed. For this, one can use a rule similar to \eqref{eq:when-2d-is-valid}, and make sure that the phase difference for two subsequent discretised $w$-values, $w_A$ and $w_B$,  is less than one radian. This results in the constraint
\begin{equation} \label{eq:minimum-w-distance}
\left|\left(w_A - w_B\right) 2\pi (\sqrt{1-l^2-m^2}-1)\right| \ll 1.
\end{equation}
This suggests that a uniform discretisation in $w$ is optimal. This is in contrast to \citet{wprojection-cornwell} where $\sqrt{w}$ tabulation is suggested. From Eq.~\eqref{eq:minimum-w-distance}, the required number of layers can be derived and is given by
\begin{equation} \label{eq:nwlayers-bound}
 N_\textrm{wlay} \gg 2\pi \left(w_{\max} - w_{\min}\right) \max_{l,m} \left(1 - \sqrt{1-l^2-m^2}\right).
\end{equation}
Actual values for the right-hand side can be very different, depending on the observation. The value of $w_{\max} - w_{\min}$ is influenced by the coplanarity of the array, the zenith angle (ZA) and the wavelength, while the value of the $\max_{l,m}$ term is influenced by the angular size of the image. For the MWA, a typical value of $w_{\max} - w_{\min}$ is $\sim 10$ at zenith, and reaches $\sim 400$ at a ZA of $30^{\circ}$. For a typical full-field-of-view image of a MWA observation of $3072\times 3072$ pixels of $0.75'$ size, the $\max_{l,m}$ term is $0.68$. This implies that tens of $w$-layers are required at zenith and hundreds at lower elevations. The number of $w$-layers has a large effect on the performance of the $w$-stacking algorithm, and will be discussed further in the next sections.

To grid the visibilities, the $w$-values are rounded to the $w$-value of the nearest $w$-layer. This discretisation can cause noticeable aliasing when using too few $w$-layers, and results in decorrelation of sources far from the phase centre in the longer baselines. Additionally, this $w$-aliasing can cause ghost sources to appear in the image. An (extreme) example of this effect is shown in Fig.~\ref{fig:aliasing-example}. When Cotton-Schwab cleaning includes prediction with too few $w$-layers, incorrect values will be subtracted from the visibilities even when no aliased sources are cleaned during minor iterations. Therefore, accurate prediction is more important than accurate imaging, because aliasing artefacts are attenuated by cleaning as long as the model is subtracted accurately.

\subsection{Computational complexity of $w$-stacking} \label{sec:time-complexity}
\begin{table}
 \caption{Scaling of the computational cost for various imaging steps, with $N_\textrm{wlay}$ the number of $w$-layers, $N_\textrm{pix}$ the number of pixels along each side, $N_\textrm{vis}$ the number of visibilities, $N_\textrm{kern}$ the size of the anti-aliasing kernel, $N_\textrm{wkern}$ the size of the $w$-kernel, $w_{\max}$ the maximum $w$ value and $\alpha_\textrm{FOV}$ the imaging FOV.} \label{tbl:computational-cost-per-operation}
 \begin{tabular}{lll}
   \textbf{Operation} & \textbf{$w$-stacking} & \textbf{$w$-projection} \\
   \hline\hline
   Fourier transform(s) & $N_\textrm{wlay} N^2_\textrm{pix} \log N_\textrm{pix}$ & $N^2_\textrm{pix} \log N_\textrm{pix}$ \\
   $w$-term corrections   & $N_\textrm{wlay} N^2_\textrm{pix}$ & $N_\textrm{vis} N^2_\textrm{wkern}$ \\
   Gridding & $N_\textrm{vis}N^2_\textrm{kern}$ & $N_\textrm{vis} N_\textrm{kern}^2$ \\
   \hline\hline
 \end{tabular}
\end{table}

It is useful to analyse the time complexity of $w$-stacking and compare it with $w$-projection, to understand which algorithm performs better in a given situation. We will use the following symbols: $N_\textrm{wlay}$ is the number of $w$-layers for $w$-stacking, $N_\textrm{pix}$ is the number of pixels in the image along each side, $N_\textrm{vis}$ is the number of visibilities, $N_\textrm{kern}$ is the size of the anti-aliasing kernel (see \S\ref{sec:gridding}), $N_\textrm{wkern}$ is the size of the $w$-kernel for $w$-projection, $w_{\max}$ is the maximum $w$ value and $\alpha_\textrm{FOV}$ is the imaging FOV.

Table~\ref{tbl:computational-cost-per-operation} shows how the computational costs scale for the operations that dominate the imaging in the $w$-stacking and $w$-projection algorithms. For comparison, we can assume the antialising kernel can be neglected and the terms $N_\textrm{wlay}$ and $N_\textrm{wkern}$ follow approximately $w_{\max} \sin \alpha_\textrm{FOV}$. The time complexity for $w$-stacking is then given by
\begin{equation}
\textrm{TC}_\textrm{wstacking}=\mathcal{O}\left(N^2_\textrm{pix} \log N_\textrm{pix} w_{\max} \sin \alpha_\textrm{FOV}+ N_\textrm{vis} \right),
\end{equation}
and for $w$-projection it is
\begin{equation}
\textrm{TC}_\textrm{wprojection}=\mathcal{O}\left(N^2_\textrm{pix} \log N_\textrm{pix} + N_\textrm{vis} w_{\max}^2 \sin^2 \alpha_\textrm{FOV}\right).
\end{equation}
From these bounds it can be concluded that in the limiting behaviour, the $w$-stacking method will be faster when the gridding of the visibilities is the dominating cost of the algorithm. The $w$-projection algorithm will be faster when the inverse FFTs are the dominant expense. In Sect.~\ref{sec:analysis} we will determine which method is faster in practice for different parameters.

\subsection{$W$-snapshot imaging} \label{sec:snapshot-imaging-theory}
$W$-snapshot imaging is a technique that combines warped-snapshot imaging with a $w$-correcting technique, such as $w$-projection or $w$-stacking \citep{widefield-imaging-ska-cornwell}. In the warped-snapshot imaging technique, the $w$-term is neglected, which results in an image with distorted coordinates \citep{perley-noncoplanar-arrays,mwa-interferometric-imaging}. Additionally, if the positions of the array elements are not perfectly planar or multiple timesteps are integrated to the same grid, visibilities will decorrelate and this can cause imaging artefacts. The original $w$-snapshot algorithm as described in \citet{widefield-imaging-ska-cornwell} corrects such artefacts by gridding visibilities on a tilted best-fit plane in $uvw$-space, and performs $w$-corrections towards that plane using $w$-projection or $w$-stacking.

We have looked into implementing snapshot imaging with $w$-corrections in a slightly different way. Instead of performing transforms of tilted planes, our method consists of phase-rotating the visibilities such that the phase centre of the observation is towards the zenith direction, i.e., the direction with minimal $w$-values. During imaging, the $w$-layers are recentred from zenith to the direction of interest by phase-shifting the visibilities, thereby translating the image over the tangent plane. This method results in $w$-corrected snapshots which need to be regridded before further integration, similar to the $w$-snapshots algorithm as described by \citet{widefield-imaging-ska-cornwell}. However, instead of creating warped images by performing the FFT over a tilted plane in $uvw$-space, this method transforms planes with constant $w$-values and produces images in zenith projection.

An image can be recentred from $(l,m)$ to $(\hat{l},\hat{m})=(l+\Delta l,m+\Delta m)$ by performing the substitution $(l,m)\rightarrow(\hat{l},\hat{m})$ in Eq.~\eqref{eq:wstacking}:
\begin{align}\notag
\frac{I'(\hat{l}, \hat{m})\left(w_{\max} - w_{\min}\right)}{\sqrt{1-\hat{l}^2-\hat{m}^2}} = \int\limits_{w_{\min}}^{w_{\max}} e^{2\pi i w(\sqrt{1-\hat{l}^2-\hat{m}^2}-1)} \times \\
\iint e^{2 \pi i \left(u\Delta l + v\Delta m\right)} V(u,v,w) e^{2\pi i \left(ul + vm\right)} du dv dw.
\end{align}
In words, recentring an image involves accounting for the position shift during the $w$-correction and final scaling, and shifting the visibilities in phase by multiplication with $e^{2 \pi i \left(u\Delta l + v\Delta m\right)}$ prior to the imaging. By doing the inverse corrections during prediction, a recentred visibility set can be cleaned with Cotton-Schwab iterations similar to a non-recentred set.

Our main reason for developing this method is that it is easier to implement, because no changes are required to the performance-critical gridding step. Another benefit of our method is that the resulting image has a circular synthesised beam, i.e., the resolution in $l$ and $m$ directions matches the intrinsic resolution of the instrument, which is desirable for cleaning. Regridding a recentred image is also more straightforward compared to regridding a warped snapshot, because in the latter case there is no analytic solution to the coordinate conversion \citep{perley-noncoplanar-arrays}. A benefit of warped snapshots is that the $w$-term errors are zero at image centre and get worse with the distance from image centre. With our approach, the effect gets worse with the distance from zenith. Our technique has a similar computational cost compared to the $w$-snapshot algorithm, although the required number of $w$-layers has a different dependency on ZA, FOV and the non-coplanarity.

Another method to shift an image is by using the periodicity of the FFT function. Since sources outside the FOV will be aliased back into the field, the image can be transformed to have the alias in the centre. This complicates gridding during imaging, because the anti-aliasing kernel needs to have a pass-band shape instead of a low-pass shape. Therefore, we chose to implement the former method of recentring the phase centre before imaging.

\section{The \textsc{wsclean} imager implementation} \label{sec:implementation}
With the purpose of testing new algorithms for the imaging of MWA data, we have written a new imager around the $w$-stacking algorithm. The imager is not specialised for the MWA, and has been successfully used for imaging VLA and GMRT data.

The new imager, called ``\textsc{wsclean}'', is written in the C++ language. It reads visibilities from \textsc{casa} measurement sets and writes output images to Flexible Image Transport System (FITS) files. Several steps are multi-threaded using the threading module of the C++11 standard library. These are: reading and writing; gridding from and to different $w$-layers; performing the FFTs; and H\"ogbom Clean iterations (peak-finding and image subtraction, \citealt{hogbom-clean}). For the latter, intrinsics are used as well. Because of these optimisations, minor Cleaning iterations with an image size of $3072\times3072$ are performed at a rate of hundreds per second. This is fast enough to make the Clark Clean optimisation \citep{clark-clean}, which consists of considering only a subset of pixels with a reduced point-spread function (PSF), less relevant. For very large images this optimisation might still be useful, but we have not implemented it in \textsc{wsclean}. Because the PSF varies for imaging with non-zero $w$-values, subtracting a constant PSF in image space leads to inaccuracies. After a number of minor iterations, it is therefore beneficial to invert the model back to the visibilities via prediction, and subtract the model directly from the visibilities \citep{wprojection-cornwell}. This is similar to the Cotton-Schwab cleaning method \citep{cotton-schwab-clean}. \textsc{wsclean} allows cleaning individual polarisations, or can jointly deconvolve the polarisations. In the latter case, peak finding can be performed in the sum of squared Stokes parameters, $I^2+Q^2+U^2+V^2$, or in $pp^2+2pq \overline{(pq)} + qq^2$ space, where $p$ and $q$ are the two polarisations and $\overline{pq}=qp$ is the complex conjugate of $pq$. After a peak has been found, the PSF is subtracted from the individual polarisations with different factors.

It will not be possible to store all $w$-layers in memory when creating large images or when many $w$-layers are used. For example, our test machine, with 32 GB of memory, can store 227 $w$-layers of $3072\times3072$. In more demanding imaging configurations, the implementation performs several passes over the measurement set and will grid a subset of $w$-layers in each pass.

\subsection{Full-sky imaging}

\begin{figure*}
\begin{center}
\includegraphics[height=22.5cm]{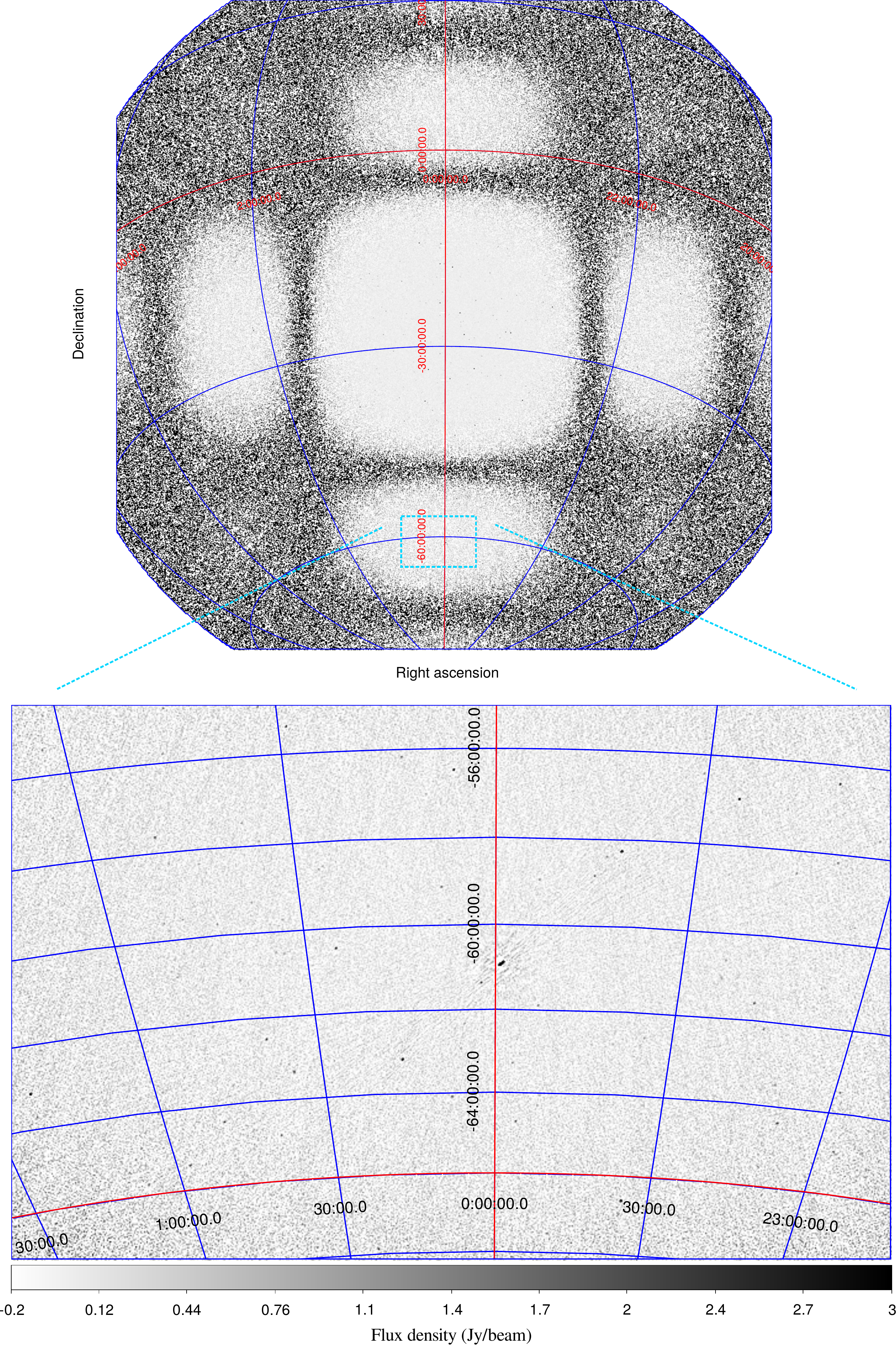}
\caption{A beam-corrected MWA image of a 112~s zenith observation at 180~MHz that covers almost the full sky. The lower image shows a zoom-in on the southern side lobe. PKS J2358-6054 ($\sim100$ Jy) is visible in the centre of the southern side lobe and resolved. The noise level in the side lobe is about 200 mJy/beam. PKS J2358-6054 has been cleaned, but some artefacts remain because no direction-dependent calibration has been performed.}
\label{fig:full-sky-example}
\end{center}
\end{figure*}

For low-frequency telescopes, it can be of interest to do full-sky (i.e., horizon to horizon) imaging. In the case of MWA, the tile beam can have strong sidelobes at a distance of more than 90\degree from the pointing centre. Imaging these sidelobes might be relevant because they are scientifically of interest (e.g. when searching for transients). Full-sky imaging can also be useful for self-calibration or deconvolution. For example, self-calibration using full-sky clean components has been found to give good results in imaging the resolved FR-II radio source Fornax~A (McKinley et al., submitted). An example of a full-sky MWA image is shown in Fig.~\ref{fig:full-sky-example}.

To efficiently image the full sky, an observation is split into short snapshots, their phase centres are changed to zenith and the snapshots are subsequently imaged, with an appropriate number of pixels and resolution. Depending on desired resolution and dynamic range, this might require very large images. To make an image at the MWA resolution, the image needs to have approximately 10k pixels along each side. The $w$-values will be very small at zenith, because at zenith only the vertical offsets of the antennas will contribute to the $w$-value. For the 128-tile MWA, the maximum differential elevation between tiles is $8.5$m. The tiles are in fact on a slight slope, and by fitting a plane to the antennas and changing the phase centre towards the normal of that plane, the maximum $w$-value decreases to $5.5\textrm{m} / \lambda$. In the following sections, when discussing the MWA zenith direction we are referring to this optimal $w$-direction.

\subsection{Implementation of $w$-snapshot imaging} \label{sec:snapshot-imaging-implementation}
\begin{figure*}
\begin{center}
\includegraphics[width=8cm]{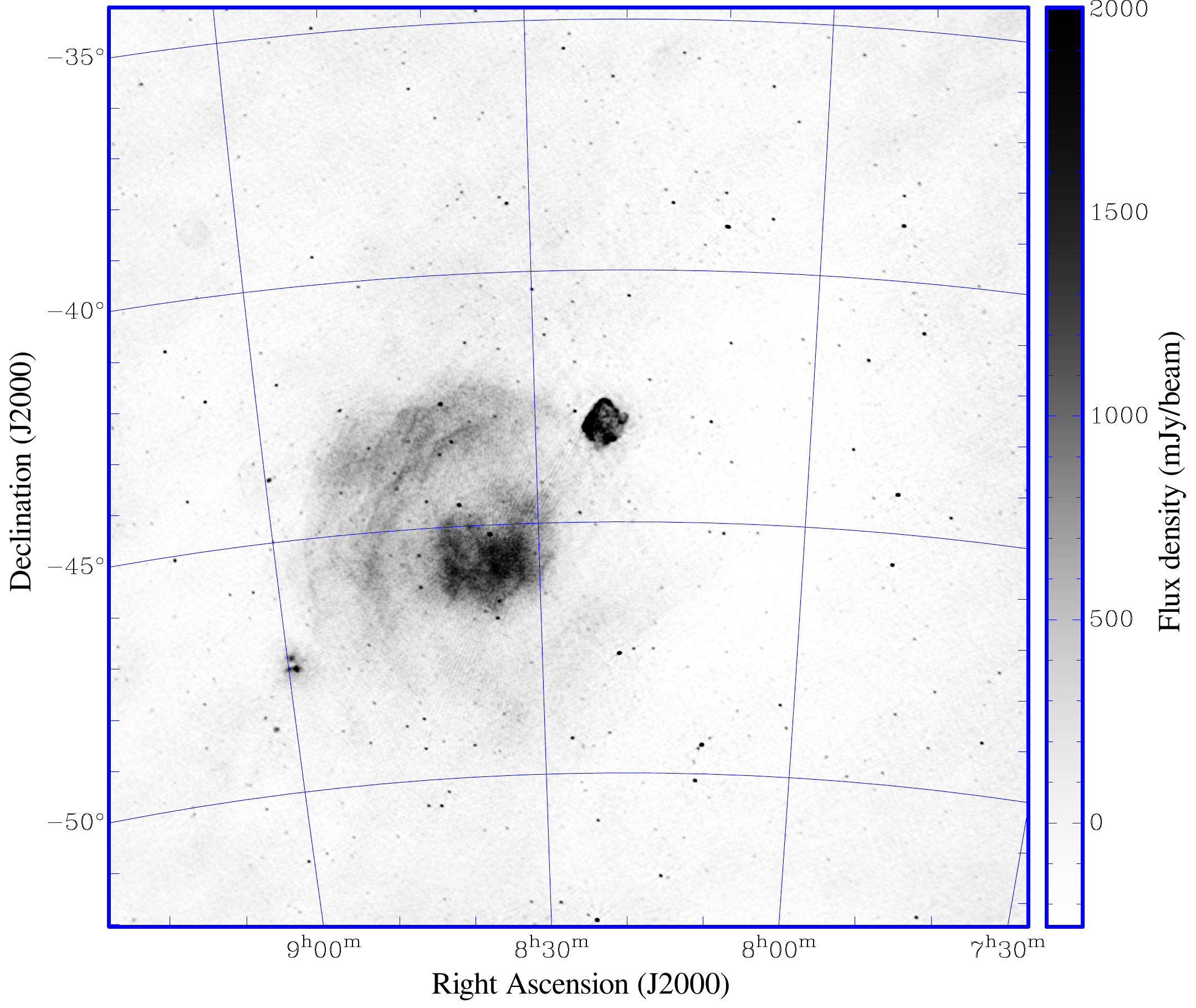}
\includegraphics[width=8cm]{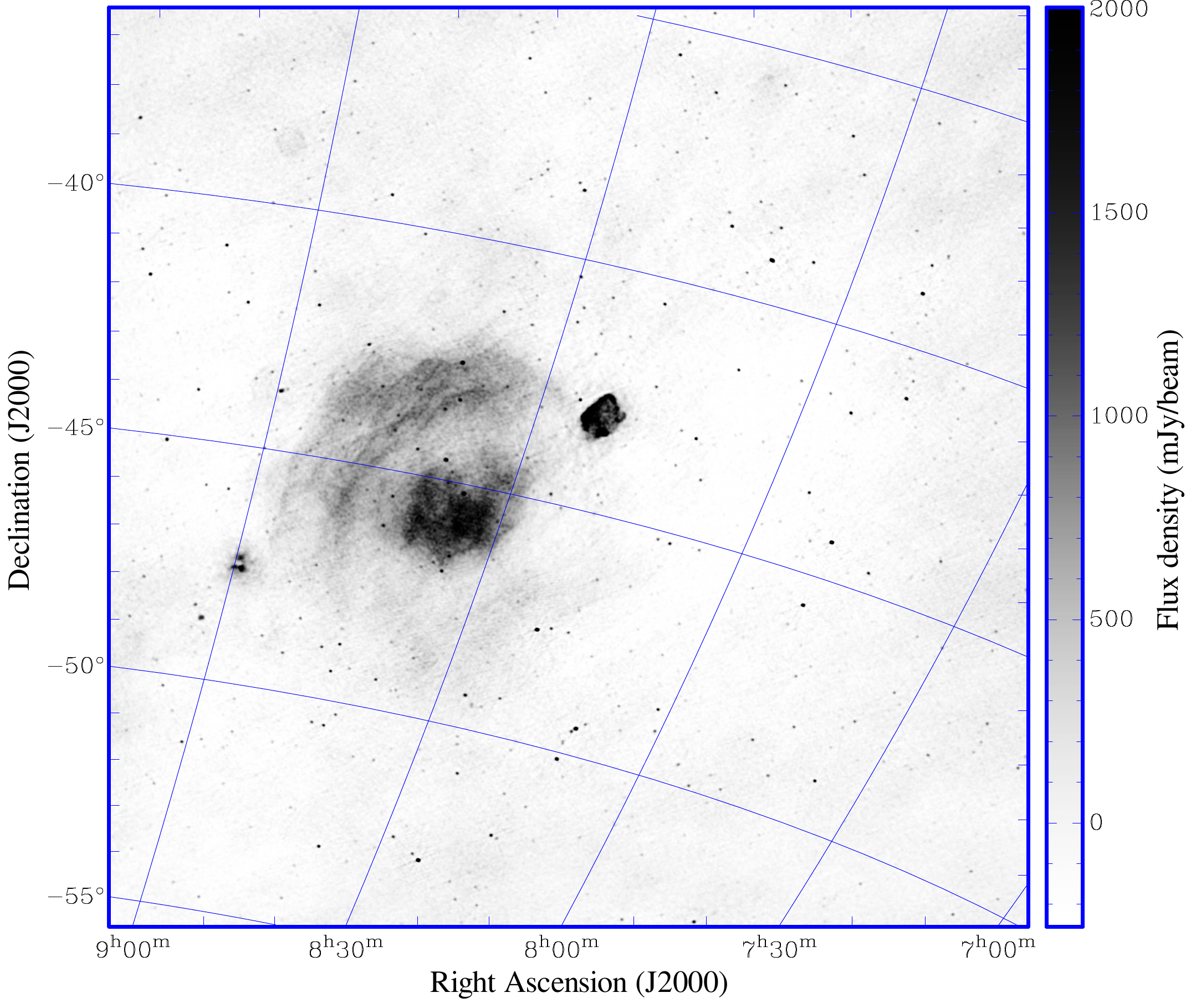}
\caption{13-min MWA observation of supernova remnants Vela and Puppis A with a centre frequency of 149~MHz. Left: normal projection centred on Puppis~A, right: phase-rotated to zenith to reduce $w$-values, recentred on Puppis A during imaging. Both Stokes-I images have been made with \textsc{wsclean} using the Cotton-Schwab clean algorithm. No beam correction was applied. Imaging computing cost was 60 min for the normal projected image and 41 min for the recentred image.}
\label{fig:vela-projection-example}
\end{center}
\end{figure*}
As discussed, all-sky snapshot imaging with zenith as phase centre is quite efficient. However, if one is not interested in imaging the whole sky, and the direction of interest is far from zenith, the computational overhead of making all-sky images is undesirable. For these cases, the recentring technique described in \S\ref{sec:snapshot-imaging-theory} was implemented in \textsc{wsclean}. This allows making smaller images with minimal $w$-values that are recentred on the direction of interest. The implementation is generic and supports interferometric data from any telescope.

A recentred image is in the projection of the zenith tangent plane and, like warped snapshots, will need an additional regridding step before multiple snapshots can be added together. A recentred image can be stored in a FITS file with the orthographic (SIN) projection normally used in interferometric imaging \citep{fits-coordinates-2002}, by setting the centre of tangent projection with the \texttt{CRPIXi} keyword \citep{wcs-in-fits}. Common viewers such as \textsc{kvis} \citep{karma-1996} and \textsc{ds9} support such FITS files and display their coordinates correctly. Unlike warped snapshots, recentred images do not need the generalised-SIN-projection keywords \texttt{PV2\_1} and \texttt{PV2\_2}.

Our implementation supports cleaning of recentred images in the same modes that normal images can be cleaned. The PSF used during minor cleaning iterations is created by multiplying the weights with the recentring corrections, such that the PSF represents a source in the middle of the image.

An example of the difference in projection between non-recentred and recentred imaging is given in Fig.~\ref{fig:vela-projection-example}, which displays a 13-min MWA observation imaged with \textsc{wsclean}. The processing steps performed for this image were: i) preprocessing of the observation with the Cotter preprocessing pipeline, which includes time averaging and RFI flagging with the \textsc{aoflagger} \citep{post-correlation-rfi-classification,scale-invariant-rank-operator}; ii) calibration using Hydra~A without direction dependence using a custom implementation of the RTS full-polarisation calibration algorithm \citep{rts-mwa}; and iii) imaging of the seven 112 s intervals separately. Cleaning an image that is in a different projection yields slightly different results, but qualitatively the two images are clearly of equal accuracy. The wall-clock time for imaging is 60 min and 41 min for normal projection and the recentred image, respectively. Of the 41 min, 3 min is spent on phase rotating the visibilities. 

\subsection{Beam correction for the MWA}
\begin{figure*}
\begin{center}
\includegraphics[width=6cm]{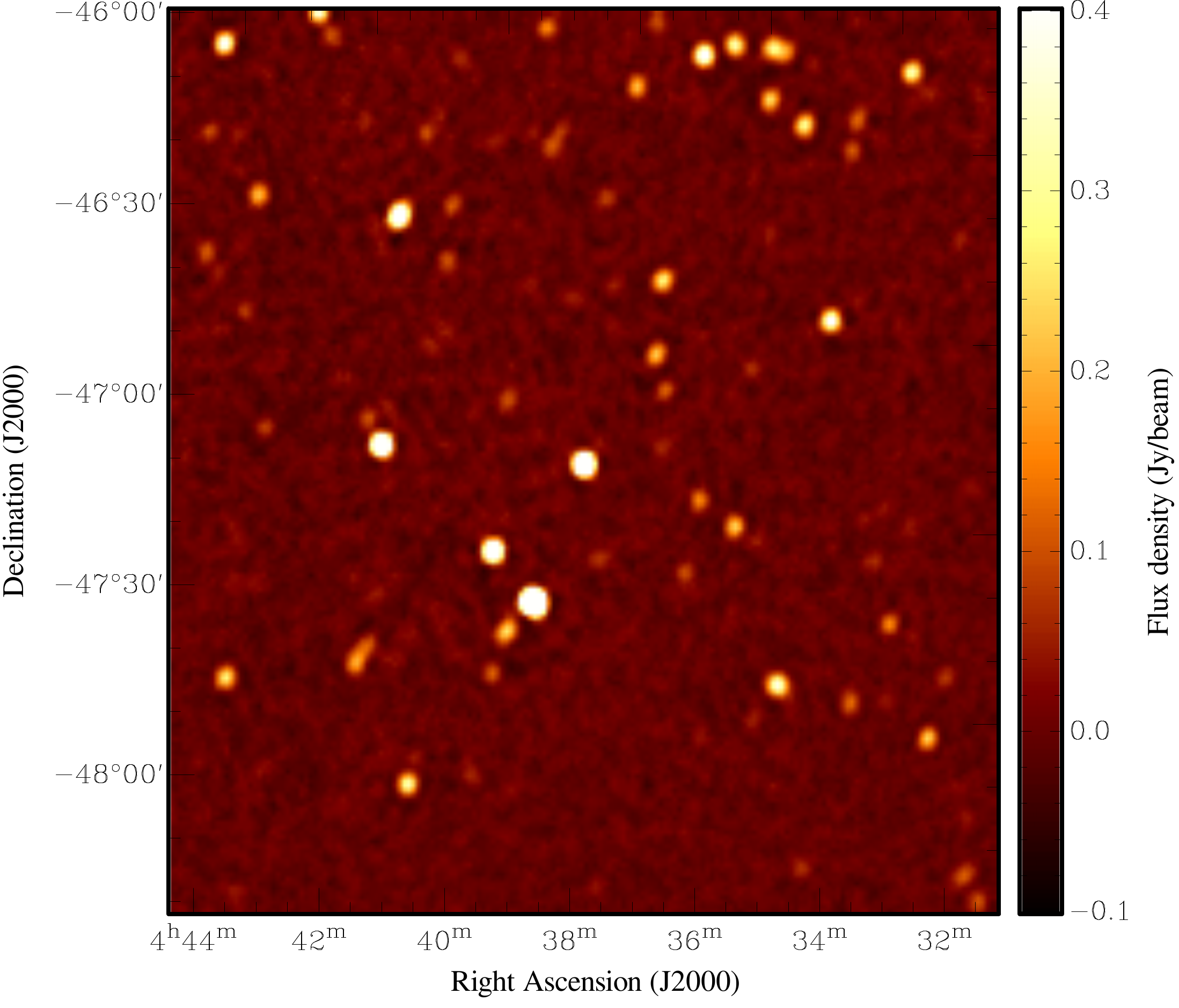}\hspace{1cm}\includegraphics[width=6cm]{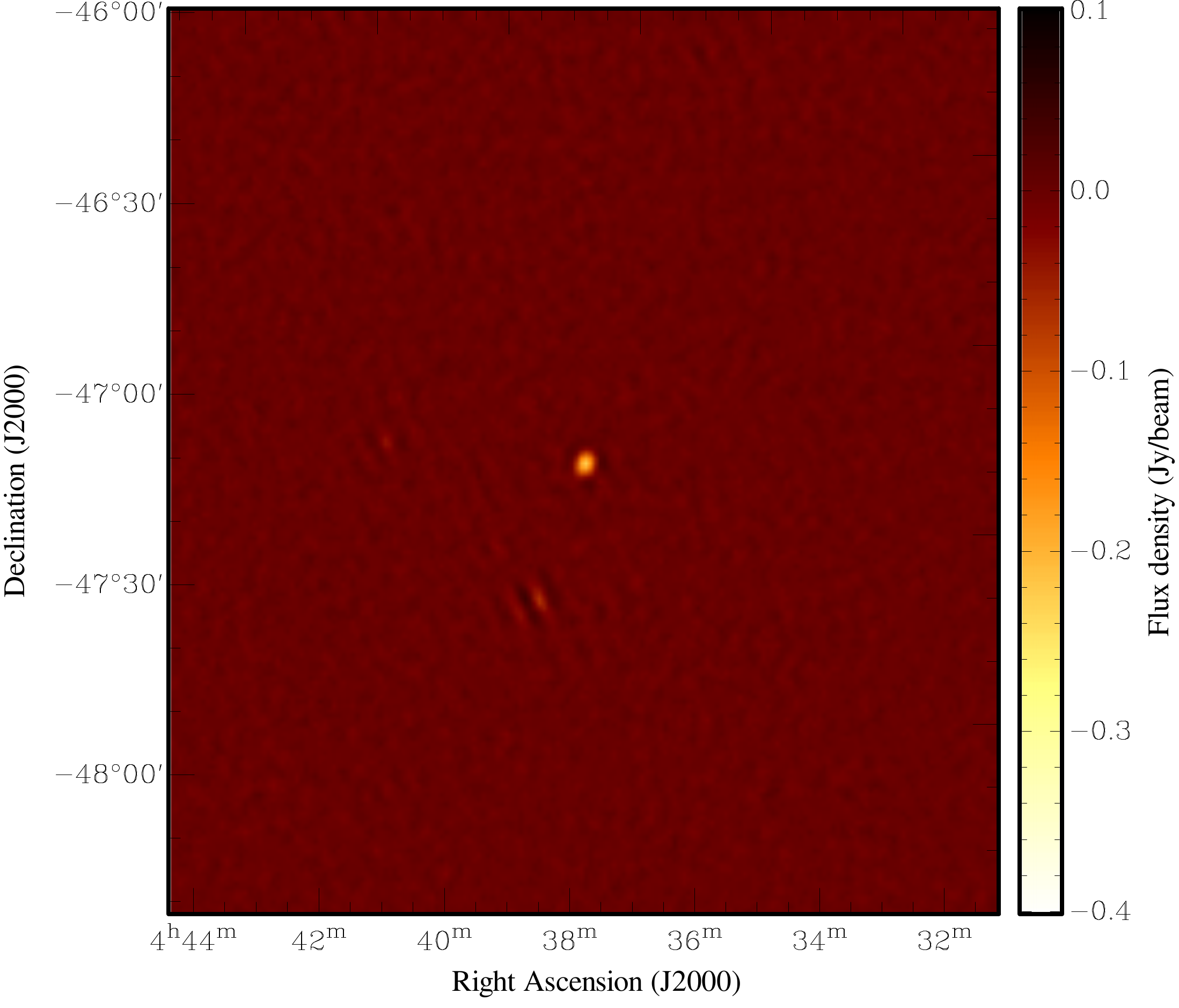}
\caption{Stokes I (left panel, total power) and Stokes V (right panel, circular polarisation) images of PSR J0437-4715, demonstrating the full-polarisation beam correction capability for MWA observations with \textsc{wsclean}. This observation was performed in drift-scan mode with a centre frequency of 154~MHz. The pulsar (centre of image) displays 17\% circular polarisation. Due to inaccuracies in the current beam model, other (unpolarised) sources show $\sim$1\% leakage into Stokes~V.}
\label{fig:pulsar-stokes-iv}
\end{center}
\end{figure*}
Because the MWA consists of fixed, beam-formed dipole antennas, the voltage beam of a MWA tile is described by a complex, non-diagonal Jones matrix. Alternatively, Mueller matrices can be used as well, but in general these result in more computations \citep{revisiting-me-i}. When assuming all individual antennas of the tiles are working properly, all tiles have the same beam. The beam can then be corrected in image space for all tiles at once with
\begin{equation}\label{eq:beam-correction}
 I(l, m) = J(l,m)^{-1} I'(l, m) \left(J(l,m)^*\right)^{-1},
\end{equation}
where each term is a $2\times2$ complex matrix, with $J$ the voltage beam and $^*$ denoting the conjugate transpose. Because $J$ is complex and non-diagonal, all four combinations of the cross-correlated polarisations\footnote{The instrumental polarisations of the MWA are informally often referred to as $X$ and $Y$, mainly because many software treats the two polarisations as such. However, this is somewhat confusing because they are not necessarily orthogonal.} $p$ and $q$ including the imaginary part of the $pq$ and $qp$ are required to calculate any of the Stokes parameters. The $pp$ and $qq$ have no imaginary part due to the $uv$-symmetry. To make proper MWA beam correction possible, \textsc{wsclean} can output both real $pp$, $pq$, $qp$, $qq$ images and imaginary $pq$ and $qp$ images. This requires four runs of the algorithm. Once these images are created, a separate program is used to perform the correction of Eq.~\eqref{eq:beam-correction}. The method is demonstrated in Fig.~\ref{fig:pulsar-stokes-iv}, which displays a detection of pulsar J0437-4715 in Stokes V. The image was made with an initial pipeline for the MWA radio-sky monitor project that uses the MWA to search for transient and variable sources. The pipeline includes \textsc{wsclean} to make the Stokes-parameter images.

Our image-based beam-correction method avoids expensive beam-correcting kernels during gridding, but requires snapshot imaging, because the MWA beam changes over time because of Earth rotation, and only works because all tiles have the same beam. To apply the same method on heterogeneous arrays such as LOFAR, each set of correlations with a different combination of station beams will have to be imaged separately, which will increase the cost of the algorithm excessively unless an a-projection kernel is used. The \textsc{awimager} uses an intermediate method, and corrects a common dipole factor in image space and the phased-array beam factor in $uv$-space, which allows the gridding kernel to be smaller compared to correcting both in $uv$-space \citep{awimager-2013}.

\subsection{Gridding} \label{sec:gridding}
A gridding convolution kernel improves the accuracy of gridding in $uv$-space \citep{optimal-gridding-schwab-1983}. A common kernel function is a windowed sinc function, which acts as a low-pass filter. This decreases the flux of sources outside the FOV, and thus helps to attenuate aliased ghost sources and sidelobes \citep{post-correlation-filtering}. By supersampling, a convolution kernel also makes it possible to place samples more accurately at their $uv$ position, thereby lowering decorrelation.

The prolate spheroidal wave function (PSWF) is generally considered to be the optimal windowing function for gridding \citep{fourier-kernel-selection-1991}. \textsc{casa}'s gridder implementation convolves samples with a PSWF of seven pixels total width during gridding. When using a variable kernel size, a PSWF is quite complicated and computationally expensive to calculate. \textsc{wsclean} currently uses a Kaiser-Bessel (KB) window function, which is easy and fast to compute, and is a good approximation of the PSWF \citep{fourier-kernel-selection-1991}.

\begin{figure}
\begin{center}
\includegraphics[width=8cm]{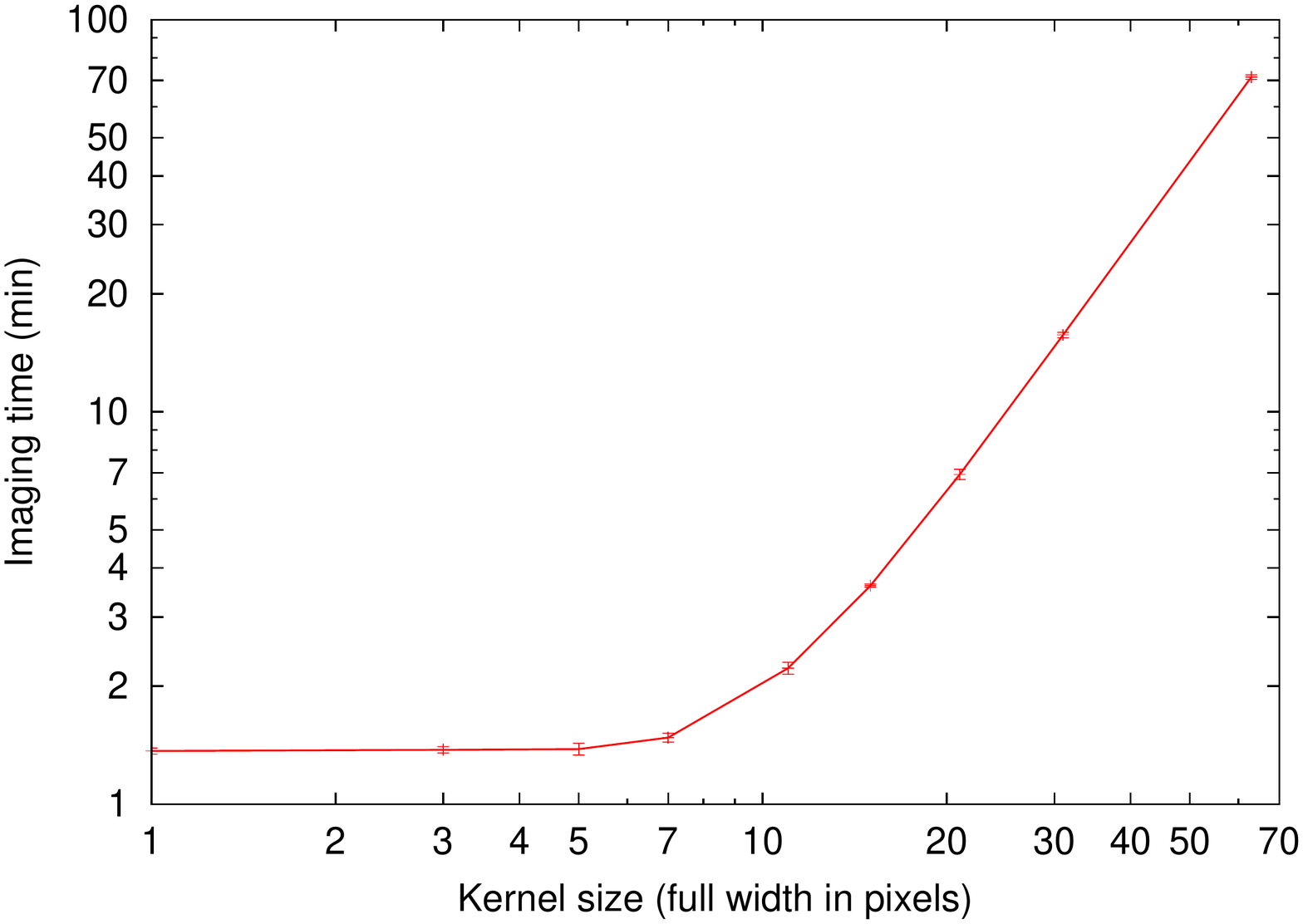}
\caption{Gridding kernel size plotted against imaging time, using common MWA settings.}
\label{fig:timing-kernelsize}
\end{center}
\end{figure}

The type of window function has no effect on the gridding performance, but the size of the kernel affects gridding performance quadratically. Fig.~\ref{fig:timing-kernelsize} shows that this quadratic relation becomes significant for kernel sizes $\gtrapprox 10$ pixels. Decreasing the kernel size to values below seven pixels has little effect on performance, because for small kernels the visibility-reading rate is lower than the gridding rate. We have not noticed much benefit of larger kernels except in rare cases where a bright source lies just outside the imaged FOV. Therefore, \textsc{wsclean} uses a default of seven pixels for the gridding kernel. It can be increased when necessary. Of course, different hardware might give slightly different results because of different reading and calculation performance.

\subsection{Lowering the resolution of inversion} \label{sec:decreasing-inversion-resolution}
Typically, for cleaning it is desired that images have a pixel size (side length) at least five times smaller than the synthesised-beam width. This improves the accuracy of the clean algorithm, because the positions of image maxima will be closer to the actual source positions. This factor of five is normally taken into account in the overall imaging resolution, i.e., the image size is increased during inversion. For the $w$-projection method, the cost of inversion with an increased resolution is small, because it does not increase the size of the $w$-kernels. It will affect the inverse FFT, but the relative cost of this step is negligible in the total cost. The $w$-stacking method is however significantly affected by the image resolution, because it performs many inverse FFTs.

The spatial frequencies in the output image are band-limited by the synthesised beam. Therefore, as long as the $uv$-plane is sampled with at least the Nyquist frequency, a high-resolution image can be perfectly reconstructed from an inversion at lower resolution. Cleaning can be performed on the high-resolution image that is reconstructed from the low-resolution image. After the high-resolution image has been cleaned, a model is created at the same (high) resolution. Because the model consists of delta functions, the spatial frequencies in the model are not band-limited. Consequently, lowering the resolution of the model image will remove information from the model. However, only the low spatial frequencies of this image will be used in the prediction, because in $uv$-space only visibilities up to the corresponding maximum baseline length will be sampled. Therefore, as long as lowering the resolution does not modify the low-frequency components, the output of the prediction step will not change.

We have implemented an option in \textsc{wsclean} to automatically decrease the inversion resolution to the Nyquist limit,
\begin{equation}\label{eq:nyquist-resolution}
 N_\textrm{Nyquist} = 2 \frac{N_\textrm{pix} S_\textrm{pix}}{S_\textrm{synth beam}},
\end{equation}
where $N_\textrm{Nyquist}$ is the resolution in pixels used during inversion, $N_\textrm{pix}$ is the requested number of pixels of the image along one side, $S_\textrm{pix}$ is the requested angular pixel scale and $S_\textrm{synth beam}$ is the minimum angular size of the synthesised beam. Before cleaning, the low-resolution image is interpolated with a procedure consisting of: i) a low-resolution FFT; ii) zero padding; and iii) a high-resolution inverse FFT. Such interpolation assumes the image to be periodic, but this is already assumed during the inversion process. After cleaning, the model is decimated by the inverse procedure, consisting of: i) a high-resolution FFT; ii) truncation; and iii) a low-resolution inverse FFT. With common settings, this procedure decreases the inversion imaging resolution by a factor of 2.5. Without further correction, such a decrease would lower the positional accuracy with which visibilities are gridded onto the $uv$-plane. This can be corrected by increasing the oversampling rate, which has almost no effect on performance.

Recreating Fig.~\ref{fig:vela-projection-example} using both the snapshot method and this optimisation lowers the computational cost from 41~min to 24~min. Of the 24~min, 77\% is spent on cleaning approximately 100,000 components. The outputs with and without lowering the inversion and prediction resolution do not visibly differ, but the difference between the residual images has an RMS of 18 mJy/beam. This can be compared to a noise level of 67 mJy/beam in the original residual image and a peak flux in the restored image of 10.3 Jy/beam. Because the difference is mostly noise like, the difference could be caused by the non-linear behaviour of clean.

\section{Analysis} \label{sec:analysis}
We will now analyse the performance and accuracy of our implementation, and compare it with the $w$-projection implementation in \textsc{casa}. For the analysis, we use imaging parameters common for MWA imaging. When a specific parameter value is not mentioned, the settings from Table~\ref{tbl:default-parameters} are used. The number of $w$-projection planes in $w$-projection is kept equal to the number of $w$-layers in $w$-stacking, and is set to the right hand side of Eq.~\eqref{eq:nwlayers-bound}. This yields 195 $w$-planes/layers at 10\degree ZA. Several configurations with large $w$-values fail to image with \textsc{casa}, because \textsc{casa} crashes during the imaging, presumably because the $w$-kernels become too large.

The software version of \textsc{casa} is ``stable release 42.0, revision 26465'', which was released September 2013. For \textsc{wsclean}, version 1.0 from February 2014 was used. The tests were run on a high-end desktop with 32 GB of memory and a 3.20-GHz Intel Core i7-3930K processor with six cores that can perform 138 giga-floating point operations per second (GFLOPS). The data are stored on a multi-disk array with five spinning hard disks, which has a combined read rate of about 450 MB/s.
\begin{table}%
\caption{Parameter values used during benchmarks, unless otherwise mentioned.} \label{tbl:default-parameters}%
\begin{center}\begin{tabular}{rl}%
\hline
Array & MWA \\
Number of elements & 128 \\
Image size & $3072 \times 3072$ \\
Angular pixel size & $0.72'$ \\
Number of visibilities & $3.5 \times 10^8$ \\
Time resolution & 2~s \\
Frequency resolution & 40~kHz \\
Observation duration & 112 s\\
Bandwidth & 30.72 MHz (768 channels)\\
Central frequency & 182 MHz \\
Zenith angle at phase centre & 10\degree \\
Max $w$-value for phase centre & 172 $\lambda$ (283 m) \\
Number of polarisations in set & 4 \\
Imaged polarisation & $pp$ ($\sim$XX) \\
Imaging mode & multi-frequency synthesis \\
Weighting & uniform \\
Data size & 18 GB \\
\hline
\end{tabular}\end{center}\end{table}

\subsection{Accuracy}
\begin{table}%
\caption{Results on imaging accuracy measurements.} \label{tbl:accuracy-measurements}%
\begin{center}\begin{tabular}{lrrr}%
\hline\hline
& \textbf{\textsc{wsclean}} & \textbf{\textsc{wsclean}}    & \textbf{\textsc{casa}} \\
&                  & \textbf{+ recentre} & \\
\hline
\multicolumn{3}{l}{\textit{Zenith angle 0\degree (12 $w$-layers/planes)}} \\
\hline
Source flux standard error & 1.31\% & & 1.34\% \\
RMS in residual image & 0.94 mJy/b & --- & 1.90 mJy/b\\
Computational time & 8.5 min & & 19.3 min\\
\hline
\multicolumn{3}{l}{\textit{Zenith angle 0\degree (128 $w$-layers/planes)}} \\
\hline
Source flux standard error & 1.39\% & & 2.08\% \\
RMS in residual image & 0.94 mJy/b & --- & 0.94 mJy/b \\
Computational time & 10.3 min & & 19.6 min \\
\hline
\multicolumn{3}{l}{\textit{Zenith angle 10\degree (195 $w$-layers/planes)}} \\
\hline
Source flux standard error & 1.75\% & 1.40\% & 2.41 \%\\
RMS in residual image & 0.90 mJy/b & 1.03 mJy/b & 1.07 mJy/b \\
Computational time & 15.3 min & 6.6 min & 178.2 min \\
\hline\hline
\end{tabular}\end{center}\end{table}

\begin{figure*}
\begin{center}
\includegraphics[height=5cm]{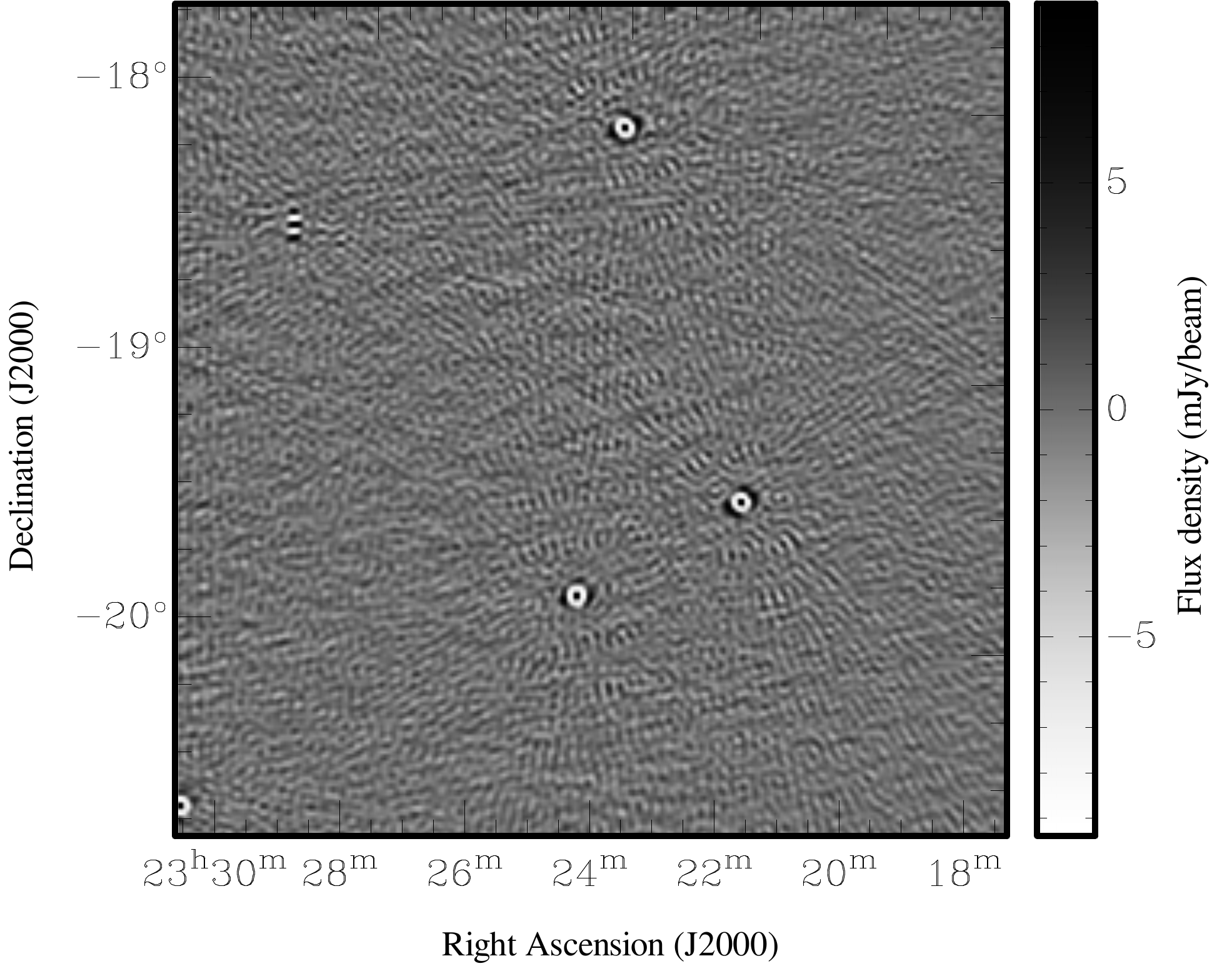}\hspace{1cm}\includegraphics[height=5cm]{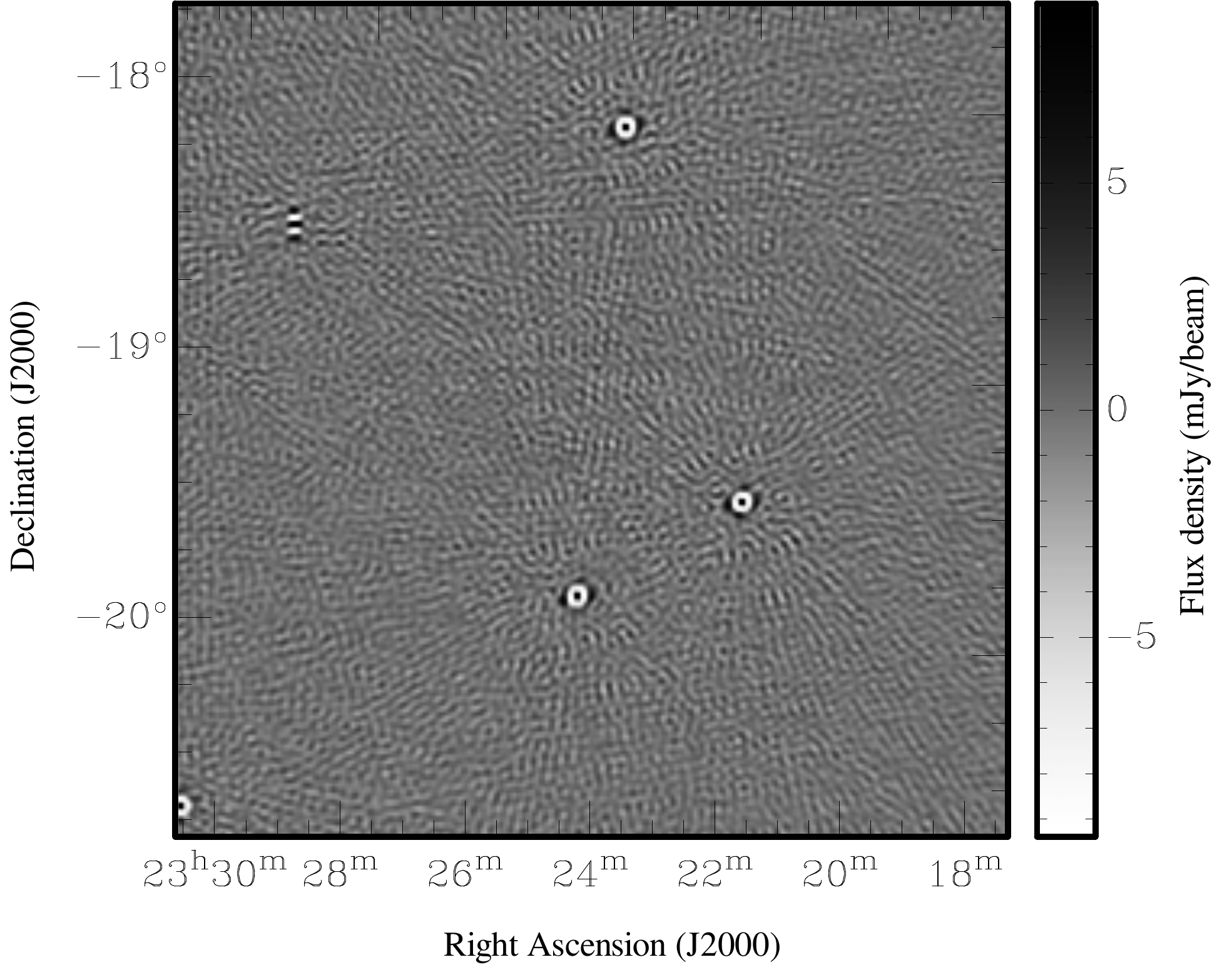}
\caption{Residual images after Cotton-Schwab cleaning of a simulated 10\degree-zenith angle MWA observation using \textsc{casa} (left) and \textsc{wsclean} (right), using similar inversion and cleaning parameters. The panels show a small part of the full images. The full field contains 100 simulated sources over 20\degree. Sources in the image produced with \textsc{casa} are slightly less accurately subtracted, leading to a residual noise level of 1.07 mJy/beam and some visible artefacts, whereas \textsc{wsclean} reaches 0.90 mJy/beam RMS noise. Since the effect is stronger away from the phase centre, it is likely that this is caused by the finite size of the $w$-kernels used in $w$-projection, which leads to inaccuracies. The ring-shaped residuals are caused by imperfect deconvolution.}
\label{fig:residuals}
\end{center}
\end{figure*}

To assess the accuracy of \textsc{wsclean} and \textsc{casa}'s clean task, we simulate a MWA observation with 100 sources of 1~Jy in a 20\degree diameter area, without adding system noise. A unitary primary beam is assumed. We image the simulated set with \textsc{wsclean} and \textsc{casa} using Cotton-Schwab cleaning to a threshold of 10~mJy. The two imagers calculate slightly different restoring (synthesised) beams, hence to avoid bias the restoring beams are fixed. Other imaging parameters are given in Table~\ref{tbl:default-parameters}. The \textsc{Aegean} program \citep{aegean-hancock-2012} is used to perform source detection on the produced images. Sidelobe noise of the residual 10~mJy source structures triggers a few false detections. These are ignored.

Table~\ref{tbl:accuracy-measurements} lists the measured root mean square (RMS) in the residual image and the standard errors of the source brightnesses as detected by \textsc{Aegean}. \textsc{wsclean} is more accurate: it shows 2-33\% lower errors in the source fluxes and produces 0-49\% lower RMS noise compared to \textsc{casa}. The large residual RMS for \textsc{casa} at zenith is caused by the fact that we keep the number of $w$-projection planes in $w$-projection equal to the number of $w$-layers in $w$-stacking, resulting in only 12 $w$-planes at zenith. This evidently has a stronger effect on the $w$-projection algorithm. However, the computational performance of the $w$-projection algorithm is hardly affected by the number of $w$-projection planes, and in practical situations one would always use more $w$-projection planes. When 128 $w$-projection planes are used in \textsc{casa}, the residual RMS is equal to \textsc{wsclean} with 12 $w$-layers, but the flux density measurements are less accurate. We do not know why this parameter needs to be higher in \textsc{casa} to reach the same RMS. We are using enough $w$-planes to cover the sources: Eq.~\eqref{eq:nwlayers-bound} results in $N_\textrm{wlay}\gg 1$ for the source furthest from the phase centre. Also unexpected is that the source flux density becomes worse by increasing the number of planes. For \textsc{wsclean} both values stay approximately the same when the number of $w$-layers is increased.

In the $\textrm{ZA}=10\degree$ case, \textsc{casa} is slightly less accurate. Extra noise can be seen in the images, as shown in Fig.~\ref{fig:residuals}. An image resulting from the technique of recentring a zenith phase-centred visibility set, as described in Sect.~\ref{sec:snapshot-imaging-theory}, is also made and analysed. As can be seen in Table~\ref{tbl:accuracy-measurements}, the source fluxes in the recentred image have smaller errors compared to the normal projection, but the residual noise is higher. The recentring technique needs on average fewer $w$-layers to reach the same level of accuracy. We have performed the tests with and without the optimisation of \S\ref{sec:decreasing-inversion-resolution}. They yield identical numbers.

\textsc{wsclean} is faster in all tested cases. Both imagers perform five major iterations for these results, and the inversions and predictions dominate the computing time. We will look more closely at the differences in performance in the next section.

\begin{figure*}%
\begin{subfigure}{.5\linewidth}%
\includegraphics[width=\linewidth]{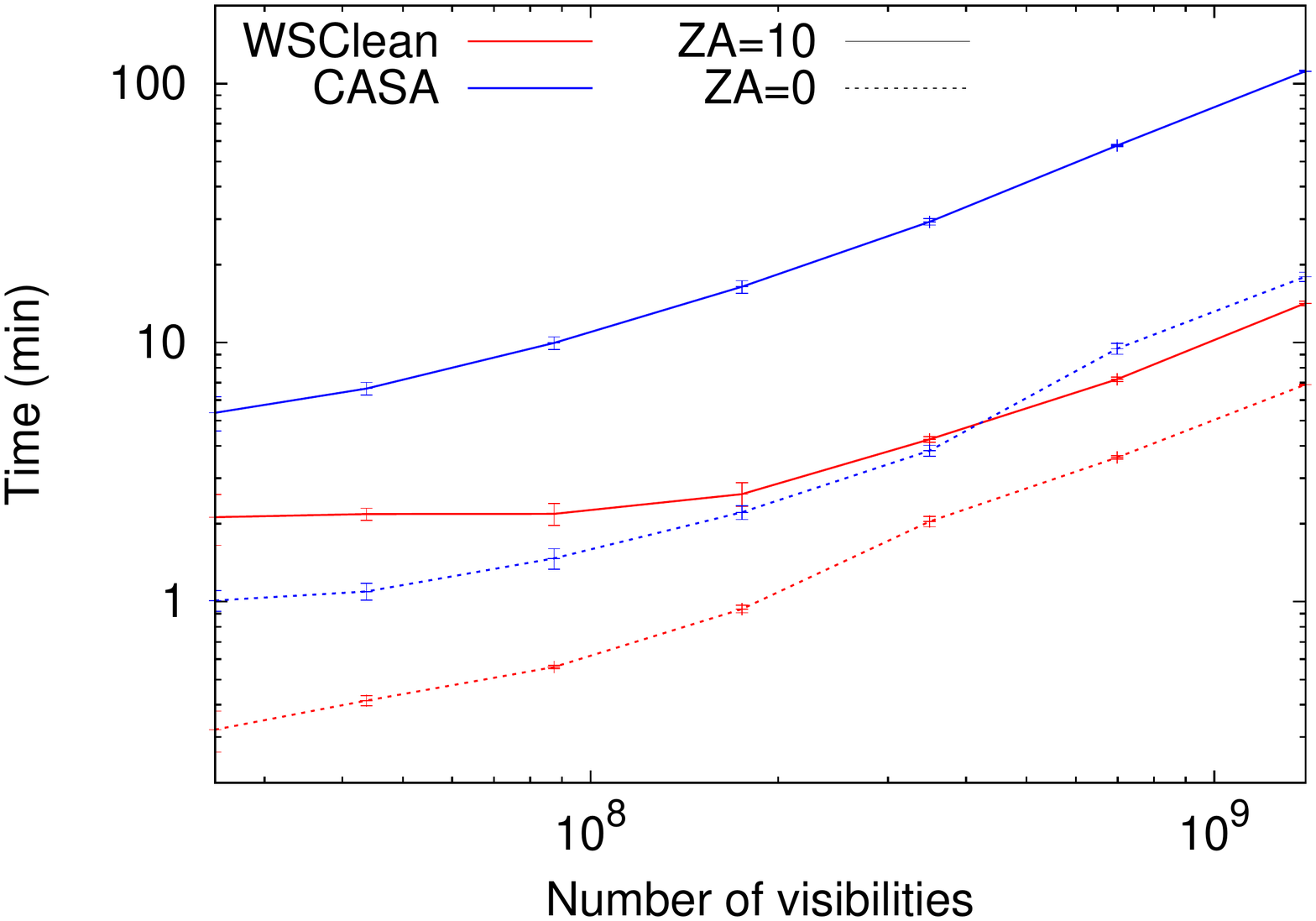}%
\caption{}\label{fig:timing-nsamples}%
\end{subfigure}%
\hspace{-.05\linewidth}\begin{subfigure}{.5\linewidth}%
\includegraphics[width=\linewidth]{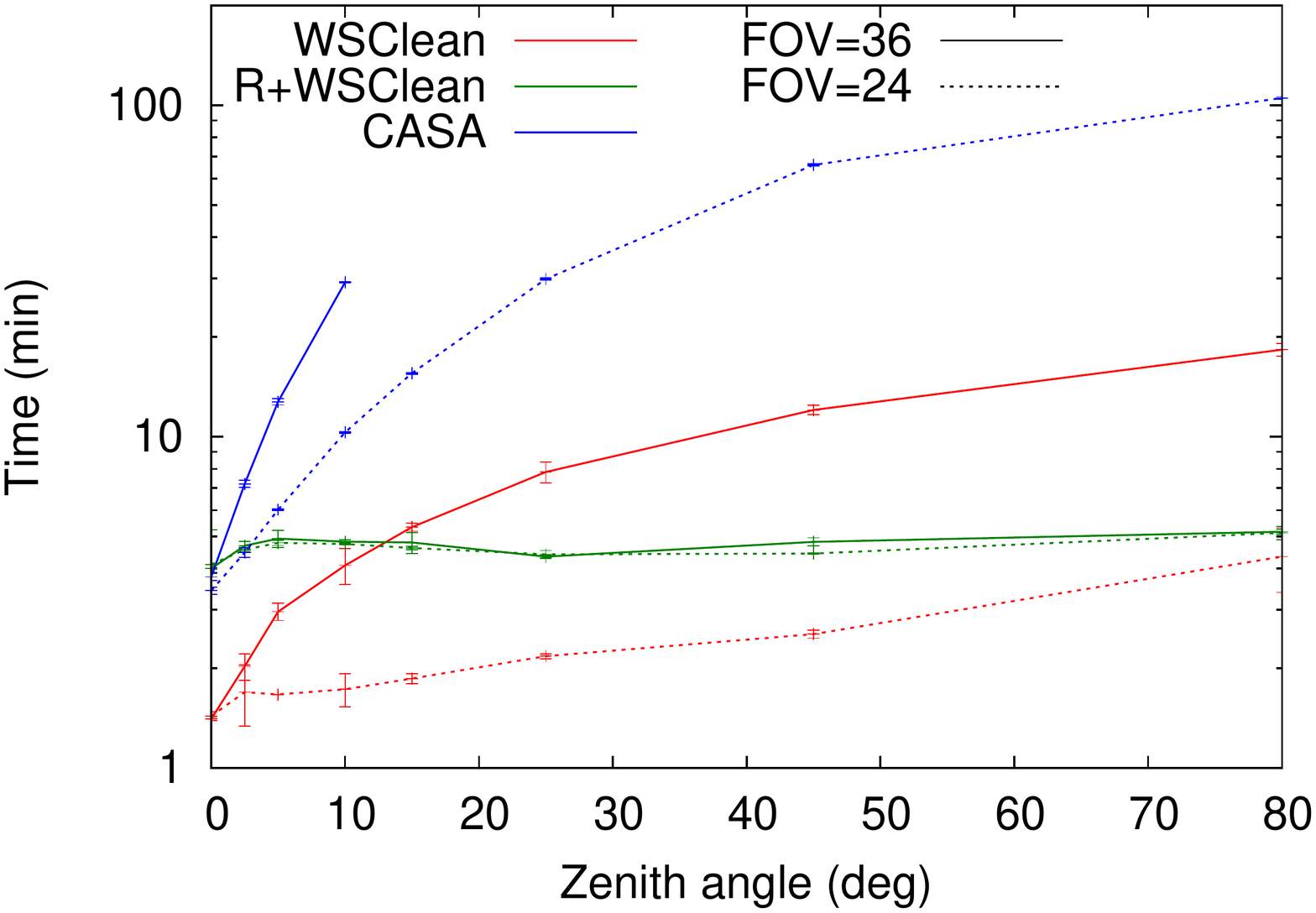}%
\caption{}\label{fig:timing-zenith-angle}%
\end{subfigure}\\
\begin{subfigure}{.5\linewidth}%
\includegraphics[width=\linewidth]{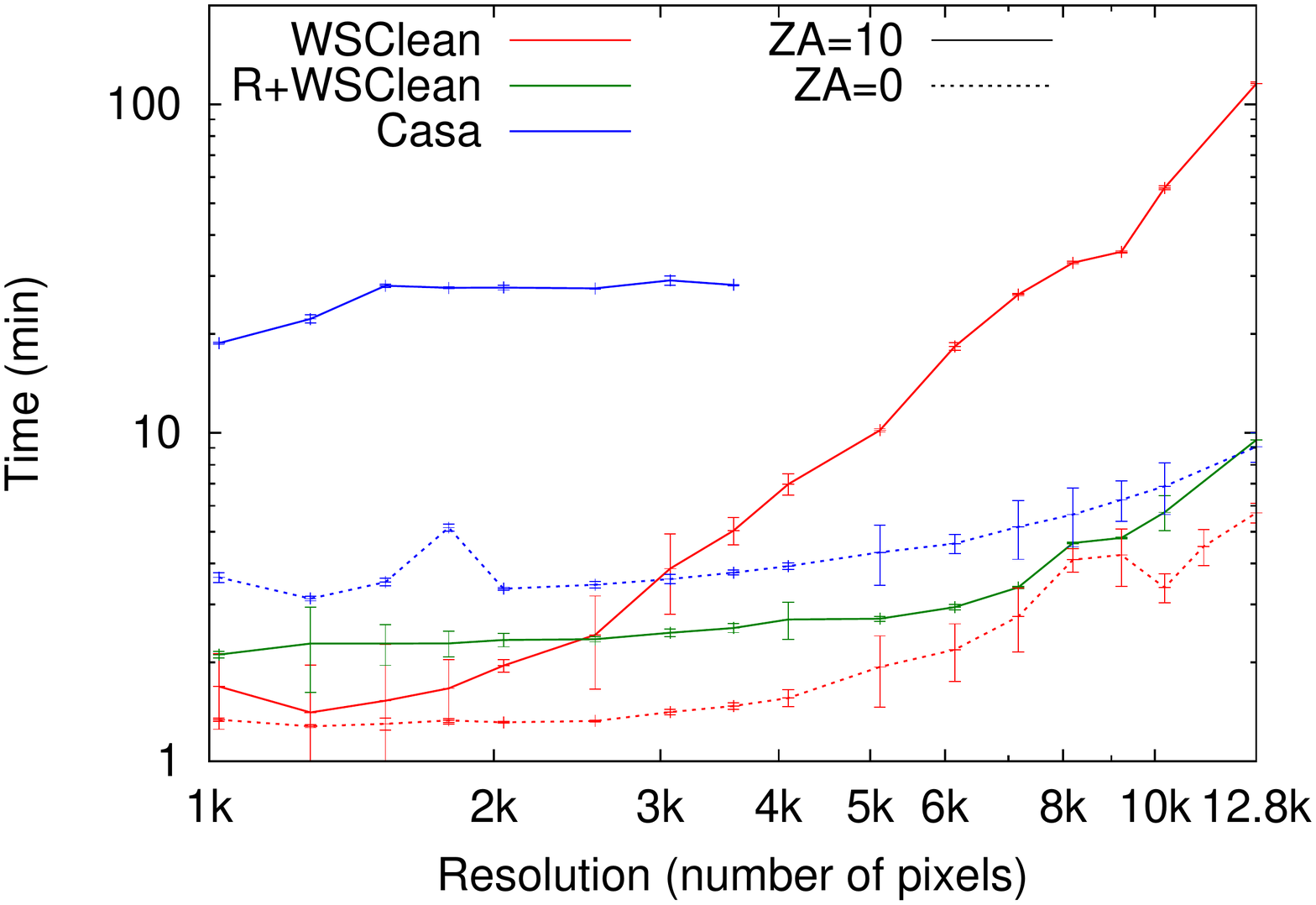}
\caption{}\label{fig:timing-resolution}%
\end{subfigure}%
\hspace{-.05\linewidth}\begin{subfigure}{.5\linewidth}%
\includegraphics[width=\linewidth]{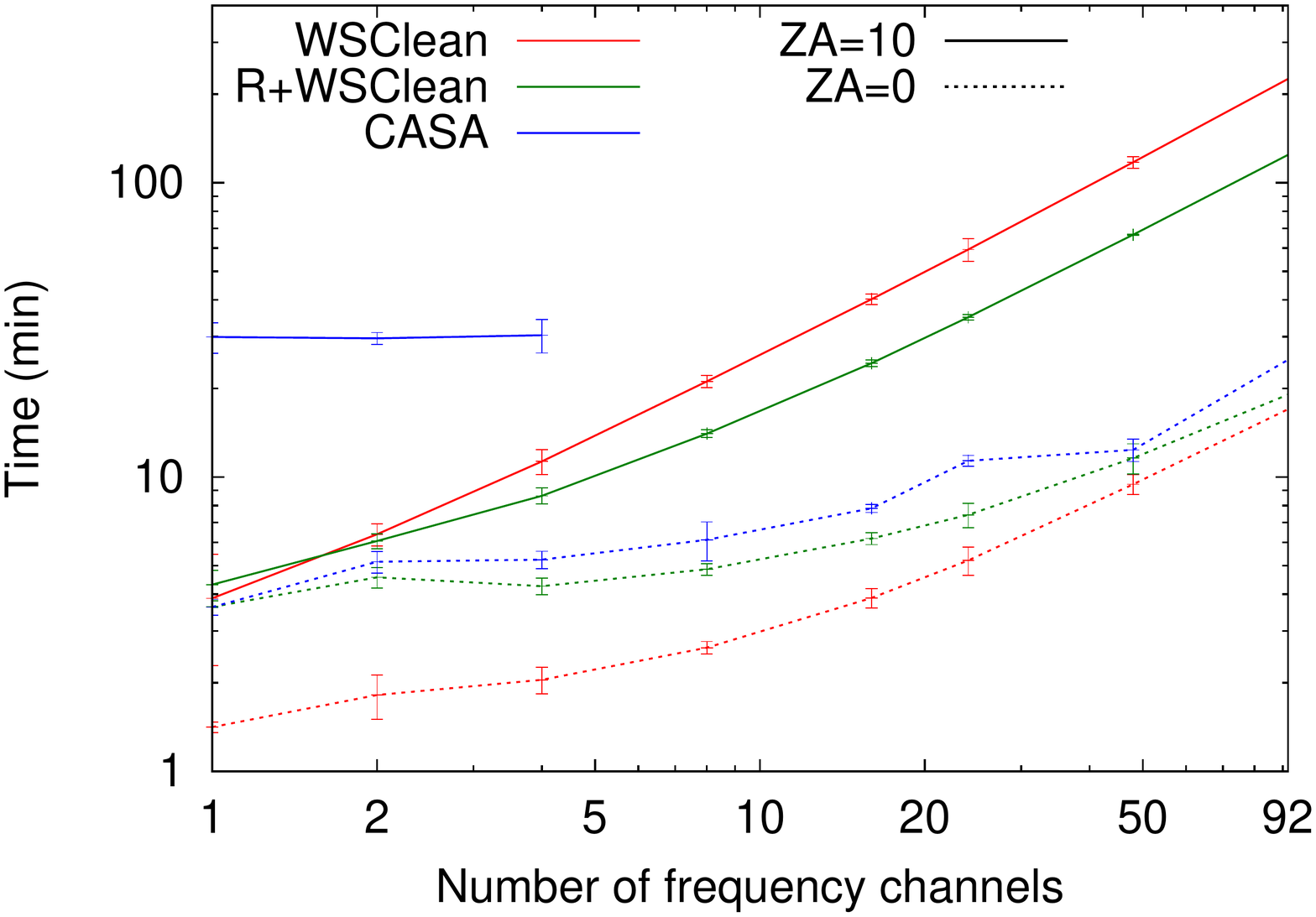}
\caption{}\label{fig:timing-channels}%
\end{subfigure}\\
\begin{subfigure}{.5\linewidth}%
\includegraphics[width=\linewidth]{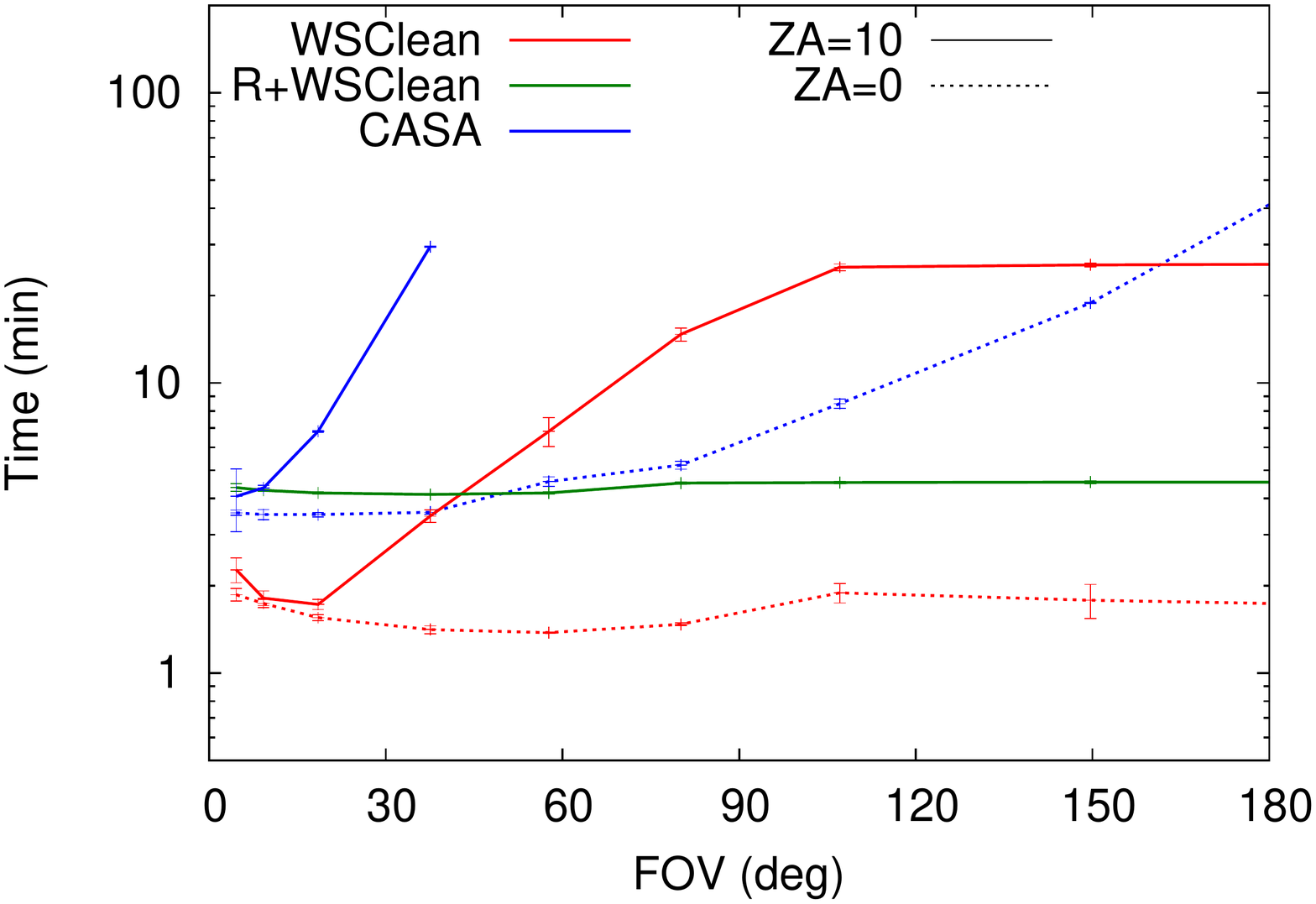}
\caption{}\label{fig:timing-fov}%
\end{subfigure}%
\caption{Imaging performance as a function of several parameters. Error bars show 5$\sigma$ level. Unfinished lines indicate the imager could not run the specific configuration successfully. The label ``R+\textsc{wsclean}'' refers to using WSClean with the recentring technique described in \S\ref{sec:snapshot-imaging-theory}.}\label{fig:timings}
\end{figure*}

\subsection{Performance analysis}

We measure the performance of the imagers using several MWA data sets. Each specific configuration is run five times and standard deviations are calculated. The variation in duration between runs is typically a few seconds. In each benchmark, the wall-clock time is measured that is required to produce the synthesised point-spread function and the image itself. No cleaning or prediction is performed, and the optimisation of \S\ref{sec:decreasing-inversion-resolution} is not used. The results are given in Fig.~\ref{fig:timings}.

Panel~\ref{fig:timing-nsamples} shows the dependency on the size of the visibility set. Results for imaging at zenith and 10\degree ZA are shown. The size of the visibility set was varied by changing the time resolution, which affects the number of visibilities to be gridded without changing the maximum $w$-value. For larger sets, both methods show a linear time dependency on the number of visibilities. This implies that gridding or reading dominates the cost. In that situation, the $w$-stacking implementation is 7.9 times faster than \textsc{casa} at $\textrm{ZA}=10\degree$ and 2.6 times faster at $\textrm{ZA}=0\degree$. With small data volumes, the FFTs start to dominate the cost, visible in Fig.~\ref{fig:timing-nsamples} as a flattening towards the left. At that point, \textsc{wsclean} is in both cases approximately three times faster. At zenith, the two methods are expected to perform almost identically. The factor of 2-3 difference in Fig.~\ref{fig:timing-nsamples} could be due to different choices in optimisations.

In Panel~\ref{fig:timing-zenith-angle}, the computational cost as a function of ZA is plotted. It shows that, as expected from \S\ref{sec:time-complexity}, compared to $w$-stacking the $w$-projection algorithm is more affected by the increased $w$-values, leading to differences of more than an order of magnitude at ZAs of $\gtrapprox 20$\degree. Additionally, at higher ZAs, the $w$-kernels become too large to be able to make images of 3072 pixels or larger. Using the recentring technique with \textsc{wsclean} to decrease the $w$-values makes the computational cost approximately constant. However, the additional time required to rephase and regrid the measurement set makes recentring only worthwhile for images $\ge3072$ pixels and $\textrm{ZA}\ge15\degree$. This is dependent on the speed of rotating the phase centre and regridding. Our code to change the phase centre is currently not multi-threaded, so its performance can be improved. Furthermore, the relative cost of changing the phase centre is lower when performing multiple major iterations.

In Panel~\ref{fig:timing-resolution}, the number of pixels in the image is changed without changing the FOV. The performance of $w$-stacking is more affected by the size of the image, but is still significantly faster in making 12.8K images at 0\degree ZA than $w$-projection. Imaging the MWA primary beam requires approximately an image size of 3072 pixels for cleaning. If the small-inversion optimisation of \S\ref{sec:decreasing-inversion-resolution} is used, the inversion can be performed at an image size of 1500 pixels. At ZA=10\degree, this saves about a factor of 2 in computing cost. The benefit of the optimisation increases with resolution: At an image size of 10,000 pixels, cost is decreased by an order of magnitude.

The cost of spectral imaging, i.e. imaging multiple frequencies, is the cost of making an image from fewer visibilities multiplied by the number of desired frequencies. Panel~\ref{fig:timing-channels} shows performance versus the spectral output resolution. As can be expected from the previous results, FFTs become the dominant cost for such small visibility sets, and the number of imaged frequencies affect performance linearly.

As can be seen in Panel~\ref{fig:timing-fov}, the FOV has a large effect on performance when imaging off-zenith. This plot was created by varying the size of a pixel in the output image, such that the number of pixels in the image did not change. The FOV is calculated as the angle subtended between the left- and right-most pixel in the image. \textsc{casa} is not able to make off-zenith images larger than $\sim$30\degree, and its imaging cost increases much more rapidly compared to the cost of \textsc{wsclean}, which implies that gridding is the major cost in this scenario. The image recentring technique is clearly beneficial for FOVs larger than $\sim$30\degree, and can even make a difference of an order of magnitude at very large FOVs of $\sim$90\degree.

\subsection{Derivation of computing cost formulae}
Based on the expected cost terms described in Sec.~\ref{sec:time-complexity}, we derive analytical functions for the \textsc{casa} and \textsc{wsclean} measurements using least-squares fitting. Several functions with different free parameters were tested, and formulae with minimum number of parameters are selected that still follow the trend of the measurements and have reasonably small errors. Measurements are weighted with the inverse standard deviation instead of the variance, because we found that the latter results in too much weight on fast configurations, leading to functions that represent the general trend less well. The single outlying measurement for $N_\textrm{pix}=1792$ with \textsc{casa} in Fig.~\ref{fig:timing-resolution} is removed.

For the \textsc{wsclean} time-cost function $t_\textrm{\textsc{wsclean}}$ we find 
\begin{align}\label{eq:wsclean-computing-cost}
& t_\textrm{WSClean} (w_{\max},N_\textrm{Mvis},N_\textrm{freq},N_\textrm{kpix}) = \\ \notag%
& N_\textrm{freq} \left( 0.526 N_\textrm{kpix}^2 \log_2 N_\textrm{kpix} (w_{\max} + 0.715) + 0.535 \right) + 0.248 N_\textrm{Mvis}
\end{align}
and for \textsc{casa} we find
\begin{align}\label{eq:casa-computing-cost}
& t_\textrm{CASA}(w_{\max},N_\textrm{vis},N_\textrm{freq},N_\textrm{pix}) = \\ \notag%
& N_\textrm{freq} \left(0.965 N_\textrm{kpix}^2 \log_2 N_\textrm{kpix} + 0.0106 N_\textrm{Mvis} (w_{\max}^2 + 40.1)\right) + 40.8.
\label{eq:casa-computing-cost}%
\end{align}
Parameter $w_{\max}$ is the maximum $w$-value and is estimated with
\begin{equation}\resizebox{0.45\textwidth}{!}{$
 w_{\max} = \frac{1}{\lambda}\left[(D \sin \textrm{ZA} + \Delta z_{\max} \cos \textrm{ZA} + \xi) \left(1.0 - \cos(\frac{1}{2}\textrm{FOV})\right)\right]$},
\end{equation}%
where $D$ is the maximum baseline length, $\Delta z_{\max}$ is the maximum height difference between antennas and $\lambda$ is the wavelength, all in meters, and $\xi$ is an extra parameter for fitting the \textsc{casa} measurements, $\xi_\textrm{CASA}=28.4$ and $\xi_\textrm{WSClean}=0$. These functions follow the trend of the measurements well, with an absolute error of 14.8\% and 20.7\% for the \textsc{wsclean} and \textsc{casa} functions, respectively.

\subsection{Optimal snapshot duration}
Snapshot imaging, such as the recentring technique described in \S\ref{sec:snapshot-imaging-theory}, can be implemented with either $w$-projection or $w$-stacking. Using the derived formulae, we can estimate the cost of making snapshots with both techniques. Snapshot imaging effectively removes the dependency on ZA from $w_{\max}$. Given the snapshot duration $\Delta \tau_\textrm{snapshot}$ and the total integration time $\Delta \tau_\textrm{total}$, the total time cost of inversion becomes
\begin{equation} \label{eq:snapshot-cost}
t_\textrm{snapshot}(\Delta \tau_\textrm{snapshot}) = \frac{\Delta \tau_\textrm{total}}{\Delta \tau_\textrm{snapshot}} t_\textrm{imaging}(w'_{\max},N'_\textrm{vis},N_\textrm{freq},N_\textrm{pix}),
\end{equation}
with
\begin{equation}
w'_{\max} = \frac{1}{\lambda}\max \Delta z \left(1.0 - \cos(\frac{1}{2}(\textrm{FOV} + \omega_E \Delta \tau_\textrm{snapshot})\right),
\end{equation}
$\omega_E$ the rotational speed of the Earth and $N'_\textrm{vis} = N_\textrm{vis}\frac{\Delta \tau_\textrm{total}}{\tau_\textrm{snapshot}}$. We have excluded the cost of phase shifting the visibilities and gridding. The cost for regridding with simple nearest neighbour or bilinear interpolation is indeed negligible, although more accurate interpolation (e.g. Lanczos interpolation) can be expensive. Also, our current implementation of the phase-changing program does take a non-negligible time, but this implementation is not optimised and can in theory be implemented in a preprocessing pipeline or on-the-fly during imaging.

The function $t_\textrm{snapshot}$ can be minimised to find the optimal snapshot duration. For \textsc{wsclean} with MWA parameters, we find that this function decreases but no minimum is reached for $\Delta \tau_\textrm{snapshot}<24$~h. Therefore, from a performance perspective, the snapshot duration should be as large as possible. In practice, the beam needs to be corrected on small time scales. A snapshot duration of more than a few minutes is therefore not possible. For $w$-projection in \textsc{casa} we find an optimal snapshot duration of $\sim2$ min for MWA observations.

\begin{table*}
\caption{Configurations for which the computational cost is predicted. Columns: $\lambda$=wavelength; FOV=field of view; Beams=number of beams; Ant=number of elements; Res=angular resolution; $D$=maximum baseline; $\max \Delta z$=maximum differential elevation; $\Delta t$=correlator dump time; BW=bandwidth; and $\Delta \nu$=correlator frequency resolution.} \label{tbl:computational-cost-configurations}
\begin{tabular}{l|rrrrrrrrrr}
 Configuration&$\lambda$& FOV  & Beams & Ant & Res & $D$ & $\max \Delta z$ & $\Delta t$ & BW & $\Delta \nu$ \\
          & (m)       & (FWHM)&       &     &     & (km)                       & (m)             & (s)    & (MHz) &  (kHz) \\
\hline
 GLEAM    &  2   & 24.7\degree& 1    & 128 & 2'   & 2.9 & 5 & 2 & 32 & 40 \\
 EMU      & 0.2  & 1\degree   & 30   &  36 & 10''  & 6 & 0.2 &10 & 300 & 20\\
 MSSS low &  5   & 9.8\degree & 5    &  20 & 100'' & 5 & 2  &10 & 16 & 16 \\
 MSSS high&  2   & 3.8\degree & 5    &  40 & 120'' & 5 & 2  &10 & 16 & 16 \\
 LOFAR LBA NL& 5 & 4.9\degree & 1    &  38 &   3'' &180& 20 & 1 & 96 & 1 \\
 AARTFAAC &  5   & 45\degree  & 1    & 288 &  20'  & 0.3& 0.5 & 1 &  7 & 24 \\
 VLSS     & 4.1  & 14\degree  & 1    &  27 &  80'' &11.1& 0.2 &10 & 1.56 & 12.2 \\
 MeerKAT  & 0.2  &  1\degree  & 1    &  64 &   6'' &  8 & 1   & 0.5 & 750  & 50 \\
 SKA1 AA core&2  &  5\degree  & 1    & 866 &    3' &  3& 5 & 10 & 250 & 1 \\
 SKA1 AA full&2  &  5\degree  & 1    & 911 &   5'' &100& 50 & 0.6& 250 & 1 \\
\end{tabular}
\end{table*}

\begin{table*}
\caption{Predicted computational costs for configurations listed in Table~\ref{tbl:computational-cost-configurations}, based on multi-frequency synthesis with five major iterations observing for 1~h at ZA=20\degree with 1 polarisation. Predictions are for $w$-projection with \textsc{casa}; $w$-stacking with \textsc{wsclean}; $w$-stacked snapshots with \textsc{wsclean} using optimal snapshot duration; and a hybrid between $w$-projection and $w$-stacking using \textsc{casa} with optimal $\Delta w$. All have approximately equal accuracy.} \label{tbl:computational-cost-predictions}
\begin{tabular}{l|rrrr|rr}
               & \multicolumn{4}{c|}{Predicted computing cost on test computer} & min & best \\
 Configuration & $w$-projection & $w$-stacking & $w$-snapshot & hybrid & computing & FLOPS/float \\
 \hline
 GLEAM    &  65h 25m & 8h 04m &  8h 03m & 14h 10m   & 1.1 TFLOPS & $6.8 \times 10^2$    \\
 EMU      &    5.2 d &  2.9 d &   2.9 d & 4.4 d     & 9.7 TFLOPS & $2.1 \times 10^4$      \\
 MSSS low &  2h 16m  & 0h 57m &  0h 57m & 2h 10m    & 130 GFLOPS & $3.4 \times 10^3$   \\
 MSSS high&  7h 16m  & 3h 52m &  3h 52m & 5h 44m    & 530 GFLOPS & $3.4 \times 10^3$   \\
 LOFAR NL &  50.2 d  &  7.3 d &   7.2 d & 12.9 d    &  24 TFLOPS & $7.2 \times 10^2$ \\
 AARTFAAC &  2.5 d   &  1.2 d &   1.2 d &   2.5 d   & 4.1 TFLOPS & $6.8 \times 10^2$   \\
 VLSS     &  0h 09m  & 0h 01m &  0h 01m &  0h 08m   & 2.2 GFLOPS & $1.1 \times 10^3$     \\
 MeerKAT  &   10.8 d &  6.2 d &   6.2 d &  10.8 d   &  21 TFLOPS & $6.8 \times 10^2$ \\
 SKA1 AA core&1581 d &  911 d &   911 d &  1570.8   & 3.0 PFLOPS & $6.8 \times 10^2$ \\
 SKA1 AA full& 643 yr& 48.8 yr & 48.8 yr& 84.1 yr   &  59 PFLOPS & $6.8 \times 10^2$ \\
\end{tabular}
\end{table*}

\subsection{Combination of $w$-projection and $w$-stacking}
To combine the $w$-projection and $w$-stacking algorithms, small $w$-corrections are made before the FFTs using a $w$-correcting kernel and large $w$-terms are corrected after the FFTs by gridding onto several $w$-layers. This allows using small $w$-kernels during the $w$-projection stage and limits at the same time the number of FFTs and required memory that pure $w$-stacking would require. The \textsc{awimager} has been applied in this way, which led to better results compared to $w$-projection without $w$-stacking \citep{awimager-2013}. For this scenario, Eq.~\eqref{eq:casa-computing-cost} can be used to determine the optimal number of $w$-layers, or more generally the optimal distance between layers, by assuming that the cost of calculating a single $w$-layer equals the cost of imaging the data set with a correspondingly smaller $w_{\max}$ value and smaller $N_\textrm{vis}$. If $\Delta w$ is the distance between $w$-layers, then 
\begin{equation}
t(\Delta w) = \frac{w_{\max}}{\Delta w} t_\textrm{CASA}(w'_{\max},N'_\textrm{vis},N_\textrm{freq},N_\textrm{pix}),
\end{equation}
with $N'_\textrm{vis} = N_\textrm{vis}\frac{\Delta w}{w_{\max}}$ and $w'_{\max} = \Delta w$. For MWA observations, the optimal value for $\Delta w$ is about unity, and improves the speed of $w$-projection by a factor of four. However, the performance of the hybrid method is still approximately a factor of two lower than our pure-stacking implementation. This could again be the difference in optimisation choices between \textsc{casa} and \textsc{wsclean}, but it does show that the effort of implementing a hybrid over pure stacking might not be worthwhile.

An optimisation suggested by \citet{awimager-2013} is to not grid visibilities with $w$-values larger than some value, because this is where most of the computational cost resides when using $w$-projection, while for LOFAR there is little benefit in gridding these samples. In $w$-stacking, the speed gain associated with this optimisation is less significant. Limiting the $w$-values also lowers the snapshot resolution in one direction significantly, because the long baselines in one direction are no longer gridded, which is often not desirable for the MWA.

\subsection{Estimated computing cost for other telescopes} \label{sec:application-to-nonmwa}
We use the derived functions to estimate the computational cost of the algorithms for configurations of several telescopes. We estimate the cost of imaging an hour of ZA=20\degree data for the following surveys or array configurations: The Galactic and Extragalactic MWA survey (GLEAM); the Evolutionary Map of the Universe (EMU) ASKAP survey \citep{emu-norris-2011}; the low and high bands of the LOFAR Multi-frequency Snapshot Sky Survey (MSSS)\footnote{See~\href{https://www.astron.nl/radio-observatory/lofar-msss/lofar-msss}{www.astron.nl/radio-observatory/lofar-msss/lofar-msss}}; all Dutch LOFAR stations \citep{lofar-2013}; the Amsterdam--ASTRON Radio Transients Facility and Analysis Centre (AARTFAAC) project\footnote{See \href{http://www.aartfaac.org}{www.aartfaac.org}}; The VLA Low-Frequency Sky Survey (VLSS, \citealt{vlss-2007}); The MeerKAT; and the low-frequency Phase~1 aperture arrays of the Square-Kilometre Array (SKA), with the full core (3~km) and the core+arms configurations \citep{ska-phase1-2013}. The configurations are summarised in Table~\ref{tbl:computational-cost-configurations}. In multi-beam configurations such as the EMU and LOFAR configurations, we assume each beam is imaged separately, i.e., the FOV of a single beam is used and the computing cost is multiplied with the number of beams. The selected wavelengths 
are approximately the central wavelength available for each instrument. The image size is set to two times the FOV width divided by the resolution, as described by Eq.~\ref{eq:nyquist-resolution}. Therefore, this assumes that the optimisation of \S\ref{sec:decreasing-inversion-resolution} to compute the inverse at lower resolution is used. The total required computing power is calculated by multiplying the estimated computing time on our test machine with the performance of the machine (138 GFLOPS).

The results are summarised in Table~\ref{tbl:computational-cost-predictions}. \textsc{wsclean} is in most situations predicted to be 2--3 times faster than \textsc{casa}. \textsc{wsclean} has the largest benefit on the full LOFAR, MWA and the full SKA, where \textsc{wsclean} is 7, 8 and 12 times faster, respectively. The $w$-snapshot method does not improve the performance much over normal \textsc{wsclean} operation in any of the cases. This might seem to differ from some of the results in Fig.~\ref{fig:timings}, where the $w$-snapshot method does show improvement in certain cases. This is because Fig.~\ref{fig:timings} tests somewhat more extreme parameters, in which the $w$-snapshot method shows more benefit. The $w$-snapshot might become more valuable at higher ZAs or when the image size needs to be larger than the half-power beam width. The hybrid method is in all situations approximately a factor of two slower than \textsc{wsclean}. All configurations listed in Table~\ref{tbl:computational-cost-configurations}, with the exception of VLSS and GLEAM, have optimal values for $\Delta w$ much smaller than one, suggesting the $w$-corrections should be done entirely by $w$-stacking instead of the $w$-stacking/projection hybrid method.

Imaging the full FOV with the full SKA becomes very expensive due to the high frequency and time resolution, and image size of $7.2\textrm{k} \times 7.2$k pixels. This translates to a computing power requirement of $\sim60$ PetaFLOPS.

\section{Conclusions} \label{sec:conclusions}
We have shown that the $w$-stacking algorithm is well-suited for imaging MWA observations. The \textsc{wsclean} $w$-stacking implementation is faster than \textsc{casa}'s $w$-projection algorithm in all common MWA imaging configurations, giving up to an order of magnitude increase in speed at a relatively small ZA of $10$\degree, and results in slightly lower imaging errors. Roughly speaking, for ZAs $>15\degree$ or FOVs $>35\degree$ snapshot imaging becomes faster. This can lead to performance improvements of a factor of $3$ for the MWA, but only in the most expensive imaging configurations that are less commonly used. Considering the extra regridding step required, which complicates issues such as calculating the integrated beam shape, recentring snapshots is worthwhile for the MWA only at very low elevations or with large fields of view. Extrapolation of the computing cost predicts that this holds for most arrays, including SKA low. A hybrid between $w$-stacking and $w$-projection does not improve performance over a pure $w$-stacking implementation, but does improve a pure $w$-projection implementation significantly. Our optimisation of lowering the image resolution during inversion and prediction increases performance by a factor of 2--10 and has no noticeable effect on the accuracy in either our test simulations, which reach approximately a dynamic range of 1:1000, or on the complicated field of Fig.~\ref{fig:vela-projection-example}.

The available SKA computing power is estimated to be around $\sim$100 PetaFLOPS. Extrapolation of our results shows that the current imagers require $3$--$60$ petaFLOPS for SKA1 low alone. Although we have measured our performance in FLOPS, it is likely that memory bandwidth will be a limiting factor, because memory bandwidth is increasing less quickly compared to the floating point performance (e.g., \citealt{gpu-gridding-romein-2012}). \citet{widefield-imaging-ska-cornwell} suggests that the $w$-snapshots algorithm improves the imaging speed for the SKA situation, but our results show that $w$-snapshot imaging does not improve SKA imaging performance over $w$-stacking. Clearly it remains challenging to perform wide-field imaging with acceptable performance, but some optimisations could be made for the SKA. For example, a performance gain of factors of a few can be achieved by averaging shorter baselines to their lowest time and bandwidth resolution before imaging. Moreover, although longer baselines make the imaging more expensive, imaging the full FOV will likely not (always) be required when using the longest baselines.

Deconvolution is not an issue for achieving low to intermediate dynamic ranges with the MWA. Because of sufficient instantaneous $uv$-coverage and near-confusion snapshot noise level, snapshots can be cleaned individually to deep levels. For example, cleaning to a $5\sigma$ level results in a cleaning threshold of approximately 100~mJy in 112~s observations. However, MWA's EoR observations \citep{bowman-science-with-the-mwa-2013} require more advanced deconvolution techniques.

So far, we have corrected the (full-polarisation) beam in image space, which is possible because of the homogeneity of the MWA tiles. This is computationally cheap for our current snapshot time of 112s, but this might be infeasible when it is required to calculate the beam on very short time scales. Accurate MWA beam models are still being developed (Sutinjo et al., in prep.). When beam corrections are required on short time scales, calculating $a$-projection kernels or performing snapshot imaging also becomes more expensive, and it would be interesting to find out where the balance lies between these methods. When only a few thousand components need to be deconvolved, an approach with direct Fourier transforms is accurate and affordable. For wide-field arrays this can be combined with calibration of the direction dependence, e.g. with the SAGECAL \citep{sage-calibration-ii} or RTS peeling \citep{rts-mwa} calibration techniques. This is not possible for fields with diffuse emission or faint point sources when no model is known \textit{a priori}.

\textsc{wsclean} is currently not able to perform multi-scale clean. Results of applying \textsc{casa}'s multi-scale clean \citep{multiscale-clean-cornwell-2008} on MWA data show that especially the Galactic plane is significantly better deconvolved using multi-scale clean, but it is very computationally expensive. Imaging a one-minute observation takes 30 hours of computational time without $w$-projection. We plan to implement some form of multi-scale cleaning in \textsc{wsclean}. It is likely that additional optimisations need to be made to be able to do this with acceptable performance. Combining multi-scale with wide-band deconvolution techniques is a possible further improvement \citep{rau-msmfs-2011}.

By using the $w$-stacking algorithms, some computational cost is transferred from gridding to performing FFTs. \textsc{wsclean} uses the FFTW library for calculating the FFTs \citep{fftw-2005}. Further performance improvement can be made by using one of the available FFT libraries that make use of graphical processing units (GPUs).

\section*{Acknowledgments}
Offringa would like to thank O.~M.~Smirnov for testing \textsc{wsclean} on VLA data.
This scientific work makes use of the Murchison Radio-astronomy Observatory, operated by CSIRO. We acknowledge the Wajarri Yamatji people as the traditional owners of the Observatory site. Support for the MWA comes from the U.S. National Science Foundation (grants AST-0457585, PHY-0835713, CAREER-0847753, and AST-0908884), the Australian Research Council (LIEF grants LE0775621 and LE0882938), the U.S. Air Force Office of Scientific Research (grant FA9550-0510247), and the Centre for All-sky Astrophysics (an Australian Research Council Centre of Excellence funded by grant CE110001020). Support is also provided by the Smithsonian Astrophysical Observatory, the MIT School of Science, the Raman Research Institute, the Australian National University, and the Victoria University of Wellington (via grant MED-E1799 from the New Zealand Ministry of Economic Development and an IBM Shared University Research Grant). The Australian Federal government provides additional support via the Commonwealth Scientific and Industrial Research Organisation (CSIRO), National Collaborative Research Infrastructure Strategy, Education Investment Fund, and the Australia India Strategic Research Fund, and Astronomy Australia Limited, under contract to Curtin University. We acknowledge the iVEC Petabyte Data Store, the Initiative in Innovative Computing and the CUDA Center for Excellence sponsored by NVIDIA at Harvard University, and the International Centre for Radio Astronomy Research (ICRAR), a Joint Venture of Curtin University and The University of Western Australia, funded by the Western Australian State government. 

\DeclareRobustCommand{\TUSSEN}[3]{#3}

\bibliographystyle{mn2e}
\bibliography{references}

\label{lastpage}

\end{document}